\documentclass{iopjournal}
\usepackage[T1]{fontenc}
\usepackage[utf8]{inputenc}
\usepackage{lmodern}
\usepackage{xurl}
\usepackage{ragged2e}
\usepackage{amsmath,amssymb,bm,braket}
\usepackage[numbers,sort&compress]{natbib}
\newcommand{\Tr}{\operatorname{Tr}}

\newcommand{\norm}[1]{\|#1\|}
\fancyhfoffset[L]{0pt}
\renewcommand{\articletype}[1]{%
  {\noindent\Large\sffamily \par}
  \vspace{5mm}{\noindent\scriptsize\sffamily\bfseries\MakeUppercase{#1}\par}
  \vspace{3mm}}
\hypersetup{pdftitle={Counterdiabatic quasi-Floquet control for the generation of entangled BICs using giant atoms},
  pdfauthor={Alexis R. Legon, Pedro Orellana, Ariel Norambuena}}
\begin{document}
\raggedbottom
\articletype{}

\title{Counterdiabatic quasi-Floquet control for the generation of entangled BICs using giant atoms}

\author{Alexis R. Leg\'on$^{1,2}$\orcid{0000-0003-1650-488X},
Pedro Orellana$^{2}$\orcid{0000-0001-7688-4111} and
Ariel Norambuena$^{2}$\orcid{0000-0001-9496-8765}}

\affil{$^1$Instituto de F\'isica, Pontificia Universidad Cat\'olica de Chile, Santiago, Chile}

\affil{$^2$Departamento de F\'isica, Universidad T\'ecnica Federico Santa Mar\'ia, Casilla 110 V, Valpara\'iso, Chile}

\email{A.R.L.: \href{mailto:alexis.legon@uc.cl}{alexis.legon@uc.cl};
\href{mailto:legon.oropeza@usm.cl}{legon.oropeza@usm.cl}\\
P.O.: \href{mailto:pedro.orellana@usm.cl}{pedro.orellana@usm.cl}\\
A.N.: \href{mailto:ariel.norambuena@usm.cl}{ariel.norambuena@usm.cl}}


\justifying

\begin{abstract}
\justifying
Bound states in the continuum (BICs) provide a mechanism for preserving
entanglement in waveguide quantum electrodynamics. Here, using quantum
control techniques in a configuration of two braided giant atoms, we
propose a robust approach to create high-fidelity entangled BICs.
The protocol combines quasi-Floquet modulation of the effective
atom--waveguide couplings with an independent phase-controlled atomic
exchange, which provides restricted counterdiabatic assistance for
preparing the complete dressed BIC. In the lossless effective model,
we obtain a full-state BIC fidelity of $0.99992$ and an unconditional
atomic Bell fidelity of $0.99588$ using a fast control protocol.
Microscopic finite-mode simulations validate the central-sideband
description during the passage, while static calibration-error scans
quantify its control tolerances. A separate microscopic study with
relaxation and dephasing, using experimentally motivated component
scales, identifies a finite useful entanglement window and a benefit
from shaping the coupling envelopes. These results support controlled
preparation, retention, and retrieval of dressed entanglement within
the stated control and noise assumptions.
\end{abstract}

\section{Introduction}

The generation and preservation of entanglement are fundamental objectives in quantum information science~\cite{Nielsen_Chuang_2010,Horodecki2009}. In waveguide quantum electrodynamics, propagating photons facilitate coherent interactions and collective decay among emitters, serving as both a resource for entanglement and a potential channel for controlling its loss~\cite{Sheremet2023}. In particular, giant atoms interact with the field at multiple, spatially separated locations. Interference between these coupling points enables control over radiative decay and exchange interactions through geometric configuration~\cite{Kockum2018,Kannan2020,drivendissipativeentanglementdistantgiant}.

Bound states in the continuum (BICs) are localized eigenstates whose energies lie within a propagating band~\cite{vN-W,Friedrich1985,ReviewNature}. In emitter-waveguide systems, they generally contain both atomic and photonic components~\cite{Facchi2016,Calajo2019BIC}. Recent studies have explored entanglement generation and dressed-state interference with giant atoms~\cite{drivendissipativeentanglementdistantgiant,Yin2023,Weng2024,Dressed.Interference.in.Giant.Superatoms}. For instance, Almanakly \emph{et al.}~\cite{drivendissipativeentanglementdistantgiant} engineered a superconducting system comprising two giant artificial atoms sequentially coupled to a common waveguide, achieving a stabilized remote Bell-state fidelity of $F = 0.89 \pm 0.02$ and a concurrence of $\mathcal C = 0.86 \pm 0.02$. These experimental results motivate the study of controlled entanglement in giant-atom systems; the calculations below address coherent preparation under a distinct set of control resources, followed by a separate dissipative feasibility study. Weng \emph{et al.}~\cite{PhysRevA.111.053711} reported high-fidelity preparation of Bell and $W$ states using BICs. Our previous work characterized geometric conditions for maximally entangled atomic components within such states~\cite{Legon2026Geometric}, motivating the dynamical preparation problem considered here.

A dynamical preparation protocol for BICs must also account for the
localized photonic field: an atomic Bell state with the waveguide in
vacuum does not, in general, constitute the complete BIC eigenstate.
Previous routes to dynamical BIC preparation include multi-photon
scattering with delayed feedback~\cite{Calajo2019BIC} and single-photon
scattering assisted by time-dependent detuning~\cite{Magnifico2025}.
Chang~\cite{Chang2026} studies coherent capture, deformation, and
release of an embedded bound state of a single giant atom, while
Guo \emph{et al.}~\cite{Guo2026BICTransfer} use time-dependent couplings
and dressed BICs for single-excitation state transfer between atomic
arrays. Building on these developments, we address the problem using counterdiabatic-assisted quantum control constructed from an effective model derived through quasi-Floquet modulation.

Quantum control is a fundamental area of study in quantum mechanics focused on manipulating the dynamics of quantum systems to optimize quantum resources~\cite{Raj2007,Dong_2010,Altafini_2012}. The field of quantum control has developed diverse methods, including spin echo~\cite{SpinEcho1950}, nuclear magnetic resonance~\cite{NMR2005}, Floquet engineering~\cite{Gandon2022,Castro_2023}, and adiabatic passage and shortcut techniques such as STIRAP~\cite{STIRAP1998,STIRAP2017}, STIREP~\cite{STIREP2022}, MOD-SATD~\cite{Baksic2016,Zhou2017}, and STA~\cite{STA2019}. Additionally, counterdiabatic strategies~\cite{SelsPolkovnikov2017,Kolodrubetz2017}, inverse engineering~\cite{InvEng1996,InvEng2012}, Lyapunov-based methodologies~\cite{Cong2013}, and machine learning algorithms~\cite{Utkan2022PINNs,NorambuenaPINNs2024} play an important role in quantum control. For entangled BIC generation, we combine two quantum control techniques: quasi-Floquet modulation to select a path within the geometrically allowed BIC family, and an independent counterdiabatic exchange that suppresses atomic-rotation errors along this selected dressed-BIC path.

In this work, we introduce a counterdiabatic quasi-Floquet protocol inspired by the coherent destruction of tunneling mechanism~\cite{Maritza2025} and counterdiabatic techniques studied in adiabatic state preparation~\cite{AlbarranArriagada2026}. First, we apply local frequency modulation to both giant atoms with slowly varying envelopes. Using Floquet theory, we derive an effective Hamiltonian that introduces a key ingredient: time-dependent atom-resonator couplings that enable coherent control within the BIC family. We then introduce a restricted counterdiabatic exchange to help prepare the selected dressed state in finite time. Together, these approaches provide an effective framework for predicting system dynamics and generating a high-fidelity, error-resilient entangled BIC. We validate the effective description against microscopic numerical simulations of the original time-dependent model, and assess the protocol's robustness by introducing errors in the control parameters.

\section{Model and effective couplings via quasi-Floquet control}

We consider two giant atoms with locally modulated transition
frequencies, coupled to a one-dimensional coupled-resonator waveguide. The dynamics is governed by the following Hamiltonian~\cite{Kockum2018,Kannan2020,Sheremet2023,Calajo2016,Wang2021Chiral,Legon2026Geometric} ($\hbar=1$)
\begin{align}\label{Hamiltonian}
H(t) &=
\sum_{i=1}^{2}\Omega_i(t)\sigma_i^{+}\sigma_i^{-}
+\sum_k\omega_k a_k^{\dagger}a_k+
\sum_k\sum_{i=1}^{2}
\left(
g_{ik}\sigma_i^{+}a_k
+
g_{ik}^{\ast}\sigma_i^{-}a_k^{\dagger}
\right).
\end{align}
Here, $\omega_k=\omega_c-2\xi\cos k$ is the bosonic dispersion relation, $\xi$ is the nearest-neighbor hopping strength of the coupled resonators, and
$g_{ik}=(2g/\sqrt{N_c})A_{k,n_i}
\exp[-ik(x_i+n_i/2)]$, where
$A_{k,n_i}=\cos(k n_i/2)$ is the geometric form factor associated
with the intra-atom separation $n_i$ and quasi-momentum
$k\in[-\pi,\pi)$, with lattice spacing set to unity. To control the atom--waveguide interaction, we introduce a local modulation
with a common carrier frequency $\nu$ and atom-dependent smooth envelopes $\varepsilon_i(t)$
\begin{align}
\Omega_i(t)
&=
\Omega_0
+
\frac{d}{dt}
\left[
\beta_i(t)\sin(\nu t)
\right]=
\Omega_0
+
\varepsilon_i(t)\cos(\nu t)
+
\frac{\dot{\varepsilon}_i(t)}{\nu}\sin(\nu t),
\end{align}
where $\Omega_0$ is the bare atomic transition frequency and
$\beta_i(t)=\varepsilon_i(t)/\nu$. The last two terms correspond to
the in-phase and quadrature components of the local modulation. This
choice ensures that the accumulated modulation phase is exactly
$\int_0^t \Omega_i(t') dt'=\Omega_0 t + \beta_i(t)\sin(\nu t)$,
even when the envelopes vary in time. The carrier is periodic, but the time-dependent envelopes generally make the complete protocol nonperiodic. We therefore use the term quasi-Floquet to denote
a modulated-envelope Floquet description~\cite{Novicenko2017}.

\begin{figure}[!ht]
\centering
\includegraphics[width=0.68\linewidth,height=0.68\textheight,keepaspectratio]{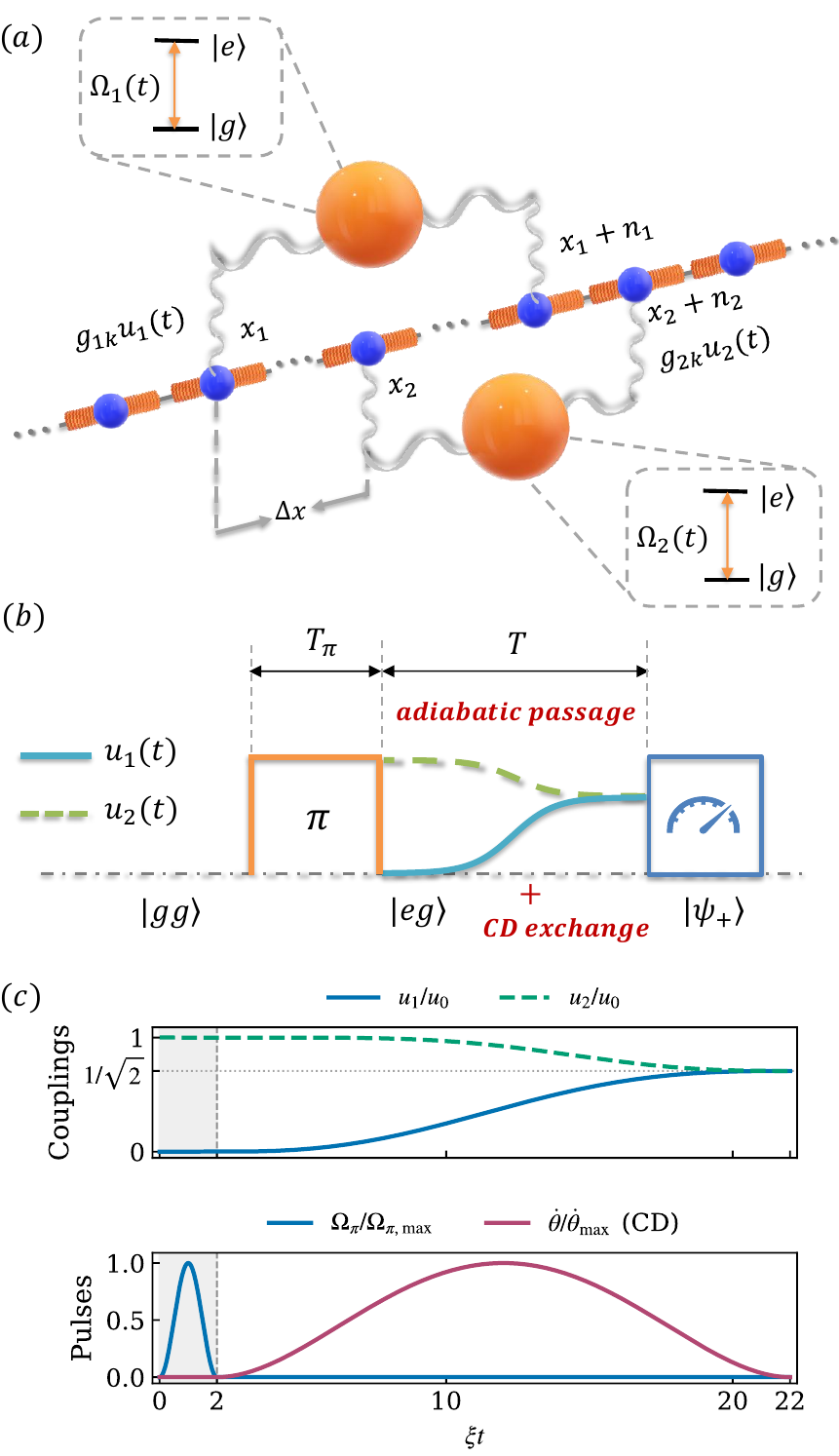}
\caption{System and control sequence for generating an entangled BIC.
(a) Two giant atoms (orange) coupled to a resonator waveguide (blue)
in a braided configuration, with nearest-neighbor hopping $\xi$.
Local quasi-Floquet modulation controls the effective couplings
$g_{ik}u_i(t)$, where $u_i(t) = \mathcal{J}_0(\beta_i(t))$ with $\beta_i(t) = \varepsilon_i(t)/\nu$. An independent phase-controlled atomic exchange
$H_{\rm cd}(t)$, not represented by the local frequency drives in
(a), supplies the auxiliary counterdiabatic (CD) control. The
numerical protocol uses connections $(0,6)$ and $(2,8)$ for giant atoms 1 and 2, respectively.
(b) Schematic of the state-preparation protocol. A local $\pi$ pulse
of duration $T_\pi$ prepares $|eg\rangle$ from the initial state
$|gg\rangle$. This is followed by a controlled coupling passage of duration
$T$, assisted by the CD exchange, which prepares the target
symmetric Bell-type dressed BIC $|B(\pi/4)\rangle$. Its normalized
atomic component is denoted $|\psi_+\rangle$ in the schematic.
(c) Temporal control sequence implementing the protocol. The upper
traces show the normalized couplings $u_1(t)/u_0$ and $u_2(t)/u_0$,
ending at $u_1=u_2=u_0/\sqrt{2}$. The lower traces show the
normalized loading pulse $\Omega_\pi(t)/\Omega_{\pi,\max}$ and the
CD exchange $\dot{\theta}(t)/\dot{\theta}_{\max}$, each normalized
to its own peak. The simulated durations are $\xi T_\pi=2$ and
$\xi T=20$, giving a total preparation time
$\xi(T_\pi+T)=22$.}
\label{fig:Figure1}
\end{figure}

We move to the interaction picture defined by
$\ket{\widetilde{\psi}(t)}=U_0^\dagger(t)\ket{\psi(t)}$, where the unitary operator is
$U_0(t)=\exp[-i\int_0^tH_{\rm local}(\tau)d\tau]$ and
\begin{equation*}
H_{\rm local}(\tau)=
\sum_i\Omega_i(\tau)\sigma_i^+\sigma_i^-
+\sum_k\omega_k a_k^\dagger a_k.
\end{equation*}
Since $[H_{\rm local}(t),H_{\rm local}(t')]=0$, no time-ordering
operator is required. In particular, using the Baker--Campbell--Hausdorff expansion, we obtain 
\begin{equation}
    U_0^\dagger(t)\sigma_i^+a_kU_0(t)
=
e^{i(\Omega_0-\omega_k)t}
e^{i\beta_i(t)\sin(\nu t)}
\sigma_i^+a_k.
\end{equation}
Thus, by applying the Jacobi--Anger identity
$\exp[i z\sin\theta]
=
\sum_{m\in \mathbb{Z}}
\mathcal{J}_m(z)e^{im\theta}$, where $\mathcal{J}_m(z)$ are the Bessel functions of the first kind, the exact interaction picture Hamiltonian becomes
\begin{align}
\widetilde{H}_{I}(t)
&=\sum_k\sum_{i=1}^{2}\sum_{m \in \mathbb{Z}} \left[g_{ik}\mathcal{J}_m[\beta_i(t)]
e^{i(\Omega_0-\omega_k+m\nu)t}\sigma_i^{+}a_k+\mathrm{H.c.}\right].
\end{align}
Although the envelopes $\beta_i(t)$ make the complete protocol
nonperiodic, the above harmonic decomposition is exact at each time.
To obtain an effective description, we follow the Floquet approach for periodically driven systems and its extension to slowly varying
envelopes~\cite{Shirley1965,Sambe1973,Goldman2014,Bukov2015,Novicenko2017}. We consider $\Omega_0$ inside the photonic band, while $\Omega_0+m\nu$ lies outside it for every relevant $m\neq0$. To retain only the central sideband, we impose the following condition throughout the protocol
\begin{align}
\Delta_F
&\equiv
\min_{\substack{m\neq0\\k \in \rm{band}}}
\left|
\Omega_0-\omega_k+m\nu
\right| 
\gg
\max_{\substack{i\\t \geq 0}}\left\{
g_{\rm col}, \left|\dot{\beta}_i\right|
\right\}.
\end{align}
Here, $g_{\rm col}=(\sum_{i,k}|g_{ik}|^2)^{1/2}=2g$ for the
connection geometry considered below; this collective scale is independent
of the momentum discretization. For $\nu>2\xi+|\delta_0|$, the bosonic dispersion gives $\Delta_F = \nu -2\xi - |\delta_0|$, where $\delta_0 = \Omega_0 - \omega_c$ is the detuning. Under these conditions, the nonzero
Floquet sidebands remain
off-resonant and the central harmonic $m=0$ determines the leading
slow dynamics. We define the local micromotion operator
$U_\mu(t)=\exp[-i\sum_j\chi_j(t)\sigma_j^+\sigma_j^-]$, with
$\chi_j(t)=\beta_j(t)\sin(\nu t)$, and
$|\Psi_{\rm slow}\rangle=U_\mu^\dagger|\Psi_{\rm lab}\rangle$.
Restoring the free evolution at $\Omega_0$ and $\omega_k$, but not the
removed local micromotion, gives the leading slow-frame Hamiltonian
\begin{align}
H_{\rm eff}(t) &=
\Omega_0\sum_{i=1}^{2}\sigma_i^+\sigma_i^-
+
\sum_k\omega_k a_k^\dagger a_k + 
\sum_{k,i}
\left[
g_{ik}^{\rm eff}(t)\sigma_i^+a_k
+ \left(g_{ik}^{\rm eff}(t) \right)^{\ast}\sigma_i^-a_k^{\dagger}\right], \\
g_{ik}^{\rm eff}(t) &=
u_i(t)g_{ik}, \qquad  u_i(t)=\mathcal{J}_0[\beta_i(t)].
\end{align}
Therefore, by varying the local modulation envelopes, we can control
the atom--waveguide couplings through the Bessel factors
$\mathcal{J}_0[\beta_i(t)]$. In particular, for
$\beta_i(t) \in [0,x_{01}]$ we have $u_i(t) \in [1,0]$, where
$\mathcal{J}_0\left(x_{01}\right)=0$ and $x_{01} = 2.4048$ is the first Bessel zero of $\mathcal{J}_0(x)$. This allows us to suppress ($u_i(t)=0$) or leave unchanged ($u_i(t) = 1$) the effective coupling by choosing the $\beta_i(t)$ profile.

Superconducting lattices have demonstrated hopping rates of
$\xi/(2\pi)=6.25\,\mathrm{MHz}$ and a frequency-control bandwidth
of $250\,\mathrm{MHz}$~\cite{Ma2019Mott}.
Using this hopping scale, a modulation frequency
$\nu/(2\pi)=80\,\mathrm{MHz}$ at $\Omega_0=\omega_c$ places all
$m\neq0$ harmonics outside the photonic band, with a minimum
detuning $\Delta_F/(2\pi)=67.5\,\mathrm{MHz}$.
Accessing the first zero of $\mathcal{J}_0(x)$ then requires
$\varepsilon/(2\pi)\simeq192\,\mathrm{MHz}$, below the
$250\,\mathrm{MHz}$ frequency excursions demonstrated
experimentally~\cite{Li2013Motional}.
These are component-level demonstrations in different devices, not
a joint validation of the proposed hardware. For smooth envelopes satisfying
$2g,\max_{i,t}|\dot\beta_i|\ll\Delta_F$,
and linewidths small compared with $\Delta_F$, only the
$m=0$ harmonic couples resonantly to the modeled photonic band;
the remaining harmonics contribute perturbative off-resonant
corrections. These experimentally motivated scales support
the central-sideband description of our protocol; finite-duration errors
are assessed numerically in Sec.~\ref{subsec:floquet_effective_validation}.
The $80\,\mathrm{MHz}$ carrier is an illustrative operating point,
distinct from the $50\,\mathrm{MHz}$ design used in
Sec.~\ref{sec:microscopic_reversible_cycle}.

\section{Instantaneous bound states in the continuum and photonic dressing}
\label{sec:bic_floquet}

To characterize the instantaneous BICs associated with the effective time-dependent couplings, we first consider the dynamics generated by $H_{\rm eff}(t)$ in the single-excitation
manifold. Since the effective Hamiltonian conserves the total number of
excitations, the general state can be written as
\begin{equation}
|\Psi(t)\rangle
=
c_1(t)|e,g,0\rangle
+
c_2(t)|g,e,0\rangle
+
\sum_k\phi_k(t)|g,g,1_k\rangle .
\label{single_excitation_state}
\end{equation}

Substituting Equation~\eqref{single_excitation_state} into the Schr\"odinger
equation $i\partial_t|\Psi(t)\rangle=H_{\rm eff}(t)|\Psi(t)\rangle$
and formally integrating the photonic amplitudes, we obtain
\begin{align}
\phi_k(t)
&=
e^{-i\omega_k t}\phi_k(0)
-i\sum_{j=1}^{2}
\int_0^t dt'\,
e^{-i\omega_k(t-t')}
\left[g_{jk}^{\rm eff}(t')\right]^{\ast}
c_j(t').
\label{phi_formal_solution}
\end{align}

Assuming that the waveguide is initially in the vacuum state,
$\phi_k(0)=0$, this gives the exact time-nonlocal equation for the giant atoms
\begin{align}
\dot c_i(t)
&=
-i\Omega_0c_i(t)
-
\sum_{j=1}^{2}
\int_0^t dt'\,
F_{ij}(t,t')c_j(t'),
\label{exact_memory_equation}
\end{align}
for $i=1,2$, where
\begin{align}
F_{ij}(t,t')
&=
\sum_k
g_{ik}^{\rm eff}(t)
\left[g_{jk}^{\rm eff}(t')\right]^{\ast}
e^{-i\omega_k(t-t')}=
\mathcal{J}_0[\beta_i(t)]
\mathcal{J}_0[\beta_j(t')]
\sum_k
g_{ik}g_{jk}^{\ast}
e^{-i\omega_k(t-t')}.
\label{exact_memory_kernel}
\end{align}
In contrast to the time-independent coupling case~\cite{Legon2026Geometric}, the kernel in
Equation~\eqref{exact_memory_kernel} depends separately on $t$ and $t'$ through the local modulation envelopes. Consequently, the memory integral does not, in general, have a convolution structure, as expected for time-dependent giant-atom couplings~\cite{Du2022Temporal}.

\subsection{Frozen resolvent and radiative cancellation}
To understand the spectral properties at a given point of the protocol, let us hold all controls fixed at a time $t_0$, treated as an independent
parameter. We then project onto the two atomic excitations and eliminate the photonic sector of this frozen Hamiltonian. In this way, we obtain
the following projected retarded resolvent
\begin{align}
G_a^R(E;t_0)
&=\big[(E+i0^+-\Omega_0)I-\Sigma^R(E;t_0)\big]^{-1},
\label{retarded_green}\\
\Sigma_{ij}^R(E;t_0)
&=\sum_k\frac{f_{ik}(t_0)f_{jk}^*(t_0)}
{E-\omega_k+i0^+}, \qquad f_{ik}(t_0)=u_i(t_0)g_{ik}.
\label{controlled_selfenergy}
\end{align}
Thus $\Sigma_{ij}^R(E;t_0)=u_i(t_0)u_j(t_0)\Sigma_{ij}^R(E)$.
We use this stationary resolvent to characterize the eigenstates at
fixed controls. For the time-dependent dynamics, we instead propagate
the finite-mode amplitude equations. This procedure retains the
two-time kernel in Equation~\eqref{exact_memory_kernel}, including the
long-time memory of the tight-binding waveguide, without freezing
the envelopes during the evolution. Using $\Sigma^R=\Delta-i\Gamma/2$, with $\Delta$ and $\Gamma$ Hermitian on the real energy axis, we obtain the decay rates
\begin{equation}
\Gamma_{ij}(E;t_0)=2\pi\sum_k f_{ik}(t_0)f_{jk}^*(t_0)
\delta(E-\omega_k).
\label{radiative_matrix}
\end{equation}
The on-shell delta functions are understood in the continuum limit.
A nonzero atomic vector $\boldsymbol{c} = (c_1, c_2)$ associated with a BIC must satisfy
\begin{align}
\sum_j f_{jk}^*(t_0)c_j&=0,\quad k=\pm k(E), \label{eq:instantaneous_dark_condition}\\
[\Omega_0 I+\Delta(E;t_0)]\boldsymbol{c}&=E\boldsymbol{c}. \label{bic_real_energy}
\end{align}
Physically, radiation must cancel in both propagation directions.
Therefore, the on-shell emission matrix must have a nontrivial
nullspace. Since nonzero real Bessel factors leave its rank unchanged,
the control amplitudes can select a dark direction only within the
conditions imposed by the connection geometry and phases. Scalar-channel reductions in the symmetric stationary limit are discussed in Appendix~\ref{app:geometric_limits}.

\subsection{Dressing and the compact BIC branch}
We write $|1\rangle=|e,g,0\rangle$, $|2\rangle=|g,e,0\rangle$,
and $|k\rangle=|g,g,1_k\rangle$. A normalized stationary dressed state can be written as
\begin{equation}
|B\rangle=\sqrt{Z}\left[
\sum_j a_j|j\rangle+
\sum_k\frac{\sum_j f_{jk}^*a_j}{E-\omega_k}|k\rangle\right],
\label{floquet_BIC_state}
\end{equation}
where the removable resonant singularities are understood by continuity
and, for $\sum_j|a_j|^2=1$,
\begin{equation}
Z^{-1}=1+\sum_k
\frac{|\sum_j f_{jk}^*a_j|^2}{(E-\omega_k)^2}.
\label{bic_normalization}
\end{equation}
A BIC satisfies the cancellation and real-energy conditions in
Eqs.~\eqref{eq:instantaneous_dark_condition} and
\eqref{bic_real_energy}, together with normalizability. Higher-order
zeros of the emission amplitude provide an additional state-selection
condition, discussed in Appendix~\ref{app:geometric_limits}. In our simulations we use $(x_1,x_1+n)=(0,6)$ and
$(x_2,x_2+n)=(2,8)$, with $\Omega_0=\omega_c$.
For connections $(0,n),(2,n+2)$, integer $n>2$, and 
\begin{equation}
u_1=u_0\sin\theta,\qquad u_2=u_0\cos\theta,\qquad q=\frac{gu_0}{\xi},
\end{equation}
an explicit compact eigenstate at $E_B=\omega_c$ is given below.
The preparation path uses $0\leq\theta\leq\pi/4$:
\begin{equation}
|B(\theta)\rangle=
\frac{|D_a(\theta),0\rangle+
(q/2)\sin(2\theta)(|1_1\rangle+|1_{n+1}\rangle)}
{\sqrt{1+(q^2/2)\sin^2(2\theta)}},
\label{compact_bic}
\end{equation}
where $|D_a(\theta)\rangle=\cos\theta|e,g\rangle+
\sin\theta|g,e\rangle$ and $|1_x\rangle$ denotes a photon at site $x$
with both atoms in their ground states. By substituting this state into the real-space hopping equations,
we verify the eigenvalue equation. The field on the two occupied
resonators cancels the four atomic sources, while all other photonic
amplitudes vanish. For $n=6$ the BIC sector at $E_B=\omega_c$ is degenerate
(zero energy in the frame rotating at $\Omega_0=\omega_c$). Two independent,
unnormalized compact states are
\begin{align}
|\mathcal B_1\rangle&=|1\rangle+
\frac{gu_1}{\xi}(|1_1\rangle-|1_3\rangle+|1_5\rangle),\\
|\mathcal B_2\rangle&=|2\rangle+
\frac{gu_2}{\xi}(|1_3\rangle-|1_5\rangle+|1_7\rangle).
\label{degenerate_bics}
\end{align}
Equation~\eqref{compact_bic} selects a particular normalized combination
within this subspace. Therefore, preparing this BIC requires control
of the selected state, rather than relying on the transport of a
unique isolated eigenstate.
The geometric amplitude relation and its extension to unequal
connection lengths are described in Appendix~\ref{app:geometric_limits},
together with numerical tests of the atomic-vacuum and frozen-envelope
approximations. Tracing over the waveguide in Equation~\eqref{compact_bic} gives
\begin{equation}
\rho_a=Z|D_a(\theta)\rangle\langle D_a(\theta)|
+(1-Z)|g,g\rangle\langle g,g|,
\end{equation}
where $Z=[1+(q^2/2)\sin^2(2\theta)]^{-1}$ is the probability
of finding the excitation in the atoms. The unconditional
observables are evaluated on $\rho_a$, without postselection.
In contrast, conditioning on one atomic excitation gives
\begin{equation}
\rho_a^{\rm cond}
=\frac{P_{1a}\rho_a P_{1a}}{Z}
=|D_a(\theta)\rangle\langle D_a(\theta)|,
\end{equation}
with $P_{1a}=|e,g\rangle\langle e,g|+|g,e\rangle\langle g,e|$.
Thus, the conditional observables characterize the normalized
atomic component, whereas the unconditional observables also
account for its weight in the complete state. The concurrence~\cite{Wootters1998} and Bell
fidelities for Equation~\eqref{compact_bic} are given by
\begin{align}
\mathcal C_{\rm cond}&=\sin(2\theta),&
\mathcal C&=Z\sin(2\theta),\\
F_{\Psi^+}^{(a)}&=\frac{1+\sin(2\theta)}{2},&
F_{\Psi^+}&=Z\frac{1+\sin(2\theta)}{2}.
\label{bic_atomic_observables}
\end{align}
Here, the subscript ``cond'' and superscript $(a)$ denote conditioning
on one atomic excitation; general dynamical definitions are given in
Appendix~\ref{app:numerics}. At $\theta=\pi/4$ the conditional atomic state is a Bell state, but the
unconditional Bell fidelity of the exact BIC is $Z<1$. We separately
measure the full-state fidelity
$F_B(t)=|\langle B[\theta(t)]|\Psi(t)\rangle|^2$.
The localized photonic dressing is part of the target BIC.
Therefore, photonic population alone cannot be identified with leakage.

Conditioning is used as a diagnostic, not as an additional
step of the preparation protocol. Its probability is $Z$ for
the exact BIC, so a high conditional fidelity alone does not
establish high unconditional preparation fidelity.
A perfectly prepared BIC has $F_B=1$, while at $\theta=\pi/4$
its unconditional atomic concurrence and Bell fidelity equal $Z$.
This reduction reflects the atom--field dressing, not a
preparation error. Conversely, atomic observables alone do not
certify the spatial profile or coherence of the photonic
component, which are also tested by the full-state fidelity.

\section{Two-stage preparation of an entangled BIC state}
\label{sec:two_stage_protocol}

We now combine local frequency modulation and two-atom exchange to
prepare the entangled state together with its localized photonic dressing.
We consider the atomic
system initially in the configuration $|g,g\rangle$, while the
coupled-resonator waveguide is assumed to be in thermal equilibrium at
sufficiently low temperature such that its Gibbs state is well
approximated by the vacuum state $|0\rangle$.

We use the basis \(\{|G\rangle,|1\rangle,|2\rangle,|k\rangle\}\), where
\(|G\rangle\equiv |g,g,0\rangle\),
\(|1\rangle\equiv |e,g,0\rangle\),
\(|2\rangle\equiv |g,e,0\rangle\), and
\(|k\rangle\equiv |g,g,1_k\rangle\). In the first stage, \(0<t<T_\pi\), a
$\pi$-pulse drive is applied only to the first atom to drive the transition $|1\rangle \leftrightarrow |G\rangle$,
\begin{equation}
H_{\pi}(t)=\frac{\Omega_{\pi}(t)}{2}
\left[
e^{i\varphi_\pi}|1\rangle\langle G|
+ e^{-i\varphi_\pi}|G\rangle\langle 1|
\right],
\label{eq:local_pi_pulse}
\end{equation}
with a smooth pulse envelope
\begin{equation}
\Omega_{\pi}(t)=\frac{2\pi}{T_\pi}\sin^2\!\left(\frac{\pi t}{T_\pi}\right),
\qquad
\int_0^{T_\pi}\Omega_\pi(t)\,dt=\pi .
\label{eq:pi_pulse_envelope}
\end{equation}
The drive phase is chosen as \(\varphi_\pi=\pi/2\), which compensates the
phase acquired under a resonant \(\pi\)-pulse and prepares
\(|G\rangle\rightarrow |1\rangle\). During this coherent driving step, the effective
couplings are held at the initial values \(u_1=0\) and
\(u_2=u_0\).
The pulse Hamiltonian is written in the frame rotating at the resonant
atomic frequency $\Omega_0$. In the ideal effective model, $u_1=0$
decouples the driven atom, and the loading dynamics remains exactly
in $\{|G\rangle,|1\rangle\}$, so no higher-excitation sector is
populated from the specified initial state. Here, we describe loading
within the ideal effective model; the microscopic validation below
is restricted to the subsequent passage.

In the second stage, \(T_\pi<t<T_\pi+T\), the pulse is switched
off and the modulation varies the effective atom--waveguide
couplings according to
\begin{equation}
u_1(\tau)=u_0\sin\theta(\tau),\qquad
u_2(\tau)=u_0\cos\theta(\tau),
\label{eq:floquet_coupling_path}
\end{equation}
where \(\tau=t-T_\pi\) is the passage time, with $\tau\in[0,T]$.
We use the smooth control function $\theta(\tau)=(\pi/4)\left(10s^3-15s^4+6s^5\right)$ with $s= \tau/T$, which satisfies the boundary conditions $\theta(0) = 0$, $\theta(T) = \pi/4$ and \(\dot{\theta}(0)=\dot{\theta}(T)=0\). For the symmetric
geometry considered here, the atomic component of the instantaneous dark
state evolves as
\begin{equation}
|D_a(\tau)\rangle =
\cos\theta(\tau)|e,g\rangle+\sin\theta(\tau)|g,e\rangle .
\label{eq:target_dark_atomic_path}
\end{equation}

The target is the selected instantaneous eigenstate $|B(t)\rangle$
of the complete central-sideband atom--waveguide Hamiltonian,
before adding auxiliary control. Because this BIC belongs to a
degenerate sector, slow evolution alone does not uniquely select
its trajectory. We derive a state-selective restricted
counterdiabatic control~\cite{Berry2009,SelsPolkovnikov2017}
by minimizing the instantaneous tracking defect over the accessible
exchange ansatz $H_a(t)=a(t)\sigma_y$:
\begin{equation}
\mathcal L(a)=\left\|(1-P_B)
\left(i|\dot B\rangle-a\sigma_y|B\rangle\right)\right\|^2,
\qquad P_B=|B\rangle\langle B|.
\end{equation}
Writing the compact target as
$|B\rangle=\cos\eta|D_a,0\rangle+\sin\eta|p\rangle$,
where $\tan\eta=(q/\sqrt2)\sin(2\theta)$ and
$|p\rangle=(|1_1\rangle+|1_{n+1}\rangle)/\sqrt2$, gives
$\mathcal L(a)=Z(\dot\theta-a)^2+\dot\eta^2$.
Its minimizer is $a=\dot\theta$, yielding
\begin{equation}
H_{\rm cd}(t)=\dot\theta(\tau)\sigma_y
=i\dot\theta(\tau)
\left(|2\rangle\langle1|-|1\rangle\langle2|\right).
\label{eq:atomic_cd_term}
\end{equation}
This exchange cancels the atomic-rotation component of the target
tracking defect, but leaves a residual of norm $|\dot\eta|$ from
the changing photonic dressing. It is therefore not an exact
counterdiabatic generator for the full BIC. The derivation of this counterdiabatic Hamiltonian is discussed in Appendix~\ref{app:restricted_cd_derivation}. In Sec.~\ref{sec:cd_residual} we analyze its action on the complete dressed BIC and compare it with an exact state-selective
auxiliary control. For our numerical simulations, we use the following finite-mode
central-sideband Hamiltonian:
\begin{align}
H_{\rm seq}(t)&=\sum_j\Delta_j|j\rangle\langle j|
+\sum_k(\omega_k-\Omega_0)|k\rangle\langle k| +\sum_{j,k}\left[u_j(t)g_{jk}|j\rangle\langle k|
+\mathrm{H.c.}\right]+H_\pi(t)+H_{\rm cd}(t).
\label{eq:two_stage_effective_hamiltonian}
\end{align}

\begin{figure}[tbp]
\centering
\includegraphics[width=\textwidth]{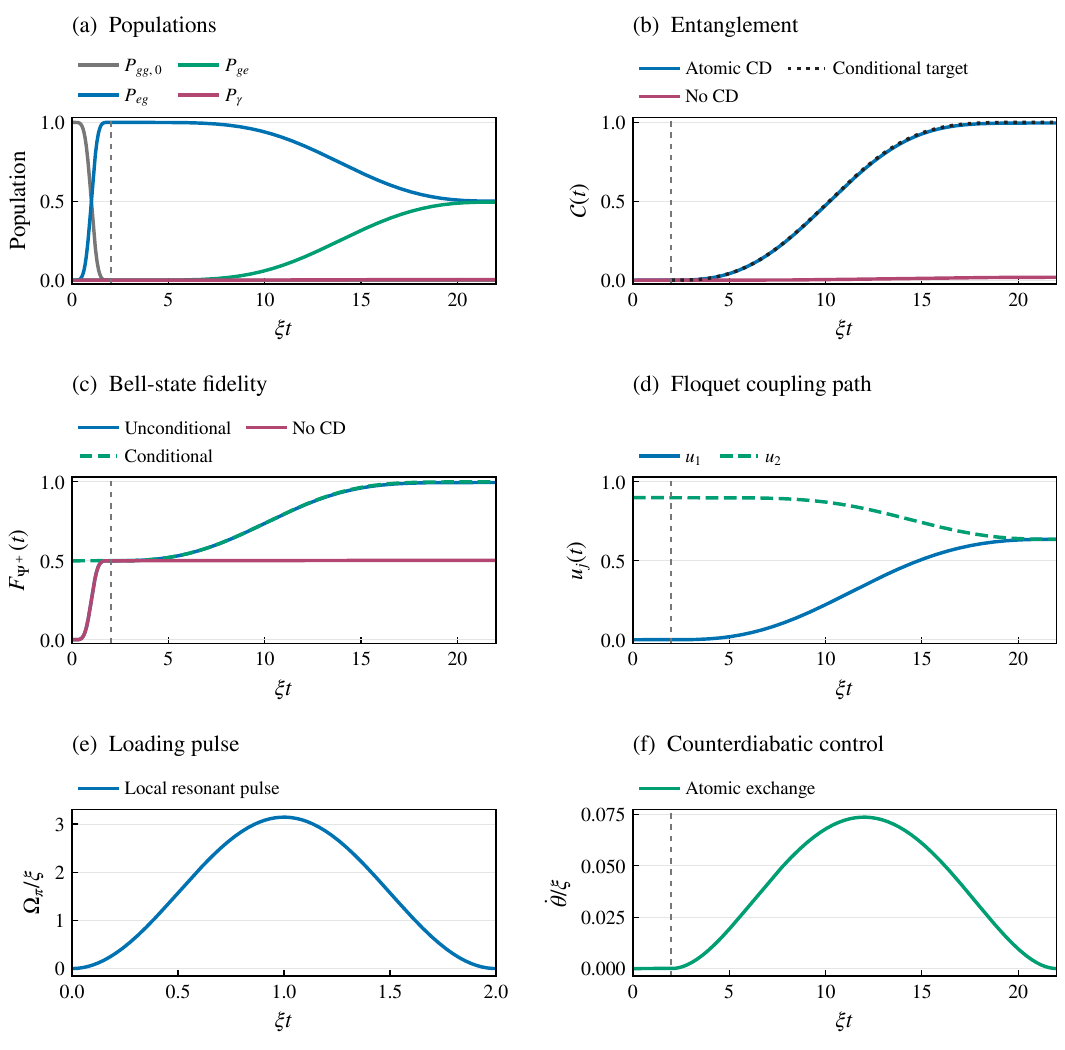}
\caption{Two-stage preparation of a Bell-type BIC.
(a) Populations during the complete coherent sequence. The curve
labeled $P_{gg,0}$ is the global-vacuum population
$|\langle g,g,0|\Psi(t)\rangle|^2$, and
$P_\gamma=\sum_k|\phi_k(t)|^2$ is the photonic population; their
sum is the reduced atomic ground-state population. A local $\pi$
pulse loads $|e,g\rangle$, followed by a CD-assisted coupling passage
that prepares the entangled atomic component and a small photonic
population. (b) Unconditional concurrence with and without atomic
CD. The curve labeled ``Conditional target'' is the conditional value
$\sin(2\theta)$; the unconditional concurrence of the exact dressed
target is $Z\sin(2\theta)$. (c) Unconditional Bell fidelity
$F_{\Psi^+}=\langle\Psi^+|\rho_a(t)|\Psi^+\rangle$ and its
conditional counterpart. (d) Effective couplings $u_1(t),u_2(t)$.
(e) Local loading pulse $\Omega_\pi(t)$, shown over the loading
interval. (f) Atomic exchange amplitude $\dot\theta(\tau)$ during
the passage. Time $t$ is measured from the start of loading, and
dashed vertical lines mark $\xi T_\pi=2$, where $\tau=t-T_\pi=0$.}
\label{fig:two_stage_bell_bic}
\end{figure}

Here, $\Delta_j$ denotes an additional atomic detuning, zero in the
nominal protocol; $H_\pi(t)=0$ during the passage and $H_{\rm cd}(t)=0$ during
the loading step. In figure~\ref{fig:two_stage_bell_bic}, we show the complete preparation sequence obtained from our numerical simulations. For \(N_c=2004\), \(\xi T_\pi=2\), \(\xi T=20\), \(u_0=0.9\), and
\(\nu/\xi=8\), the loading pulse prepares the single-excitation product state
with an error of \(|1-P_{eg}(T_\pi)| \approx 10^{-13}\). In the absence of the
counterdiabatic correction, the final concurrence remains small. By contrast,
the CD-assisted protocol reaches a final concurrence
$\mathcal{C}=0.9959$, an unconditional Bell fidelity
\(F_{\Psi^+}=0.9959\), and a conditional atomic Bell fidelity
\(F_{\Psi^+}^{(a)}=0.99999\). These results show that the CD-assisted coupling path transfers
a locally prepared excitation into the Bell component of a BIC.
In the following subsections, we analyze how well the protocol also
prepares the complete dressed state.

\subsection{Restricted control and its exact dressed reference}
\label{sec:cd_residual}
To understand the role of the coupling path, let us first consider
the auxiliary exchange without the waveguide interaction. For $g=0$,
$H_{\rm cd}=\dot\theta\,\sigma_y$ maps $|e,g\rangle$ exactly to
$\cos\theta|e,g\rangle+\sin\theta|g,e\rangle$. Thus, a pulse area
$\int_0^T\dot\theta\,dt=\pi/4$ is sufficient to generate a Bell state.
The contribution of the waveguide-coupling path must therefore be
understood through the preparation of the photonic dressing and the
subsequent retention of the state. Restricting the auxiliary control
to accessible operators is a general strategy in approximate
counterdiabatic driving~\cite{SelsPolkovnikov2017}. Here, we analyze
this restriction by calculating the residual error in following the
selected state.

For a smooth normalized selected state $|B(t)\rangle$, the state-selective
transitionless generator can be written as
\begin{equation}
H_{\rm tr}=i[\dot P_B,P_B],\qquad P_B=|B\rangle\langle B|.
\label{exact_projector_cd}
\end{equation}
To quantify the error for an auxiliary Hamiltonian $H_{\rm aux}$,
we define the tracking residual
\begin{equation}
|r(t)\rangle=(1-P_B)\big(i|\dot B\rangle-H_{\rm aux}|B\rangle\big).
\label{tracking_residual}
\end{equation}
Choosing the target phase to match the diagonal dynamical and geometric
terms, unitarity and Duhamel's identity give, for initial state $|B(0)\rangle$
\begin{equation}
1-F_B(T)\leq
\left[\int_0^T\norm{r(t)}\,dt\right]^2 .
\label{duhamel_bound}
\end{equation}
This bound applies to the unitary atom--field dynamics without
requiring a spectral gap. To evaluate it for Equation~\eqref{compact_bic},
we introduce
\begin{equation}
|p\rangle=\frac{|1_1\rangle+|1_{n+1}\rangle}{\sqrt2},\qquad
\eta=\arctan\!\left[\frac{q}{\sqrt2}\sin(2\theta)\right].
\end{equation}
Then $|B\rangle=\cos\eta\,|D_a,0\rangle+\sin\eta\,|p\rangle$.
The atomic term in Equation~\eqref{eq:atomic_cd_term} follows the rotation of
$|D_a\rangle$ but leaves the residual
\begin{align}
|r\rangle&=i\dot\eta\big(-\sin\eta\,|D_a,0\rangle
+\cos\eta\,|p\rangle\big), \qquad
\dot\eta=\frac{\sqrt2 q\cos(2\theta)}
{1+(q^2/2)\sin^2(2\theta)}\dot\theta,
\qquad \norm{r}=|\dot\eta|.
\label{explicit_tracking_residual}
\end{align}
A state-selective exact implementation adds
\begin{equation}
H_{\rm dress}=i\dot\eta\left(
|p\rangle\langle D_a,0|-|D_a,0\rangle\langle p|\right).
\label{dressed_auxiliary}
\end{equation}
The resulting $H_{\rm cd}+H_{\rm dress}$ transports the complete compact
state exactly. We use this control as a theoretical reference, since
it requires coupling to the localized photon mode and is not generated
automatically by longitudinal atomic modulation.

For the monotonic passage to $\theta=\pi/4$,
Equation~\eqref{duhamel_bound} reduces to
$1-F_B(T)\leq\arctan^2(q/\sqrt2)=0.0040391$ at $q=0.09$.
The observed error, $7.82\times10^{-5}$, is well below this conservative
bound. Changing the duration alone leaves this integrated bound unchanged.
Therefore, we study the time-dependent errors by directly propagating
the state.

\subsection{Dressed-state retention and retrieval}
\label{subsec:retention}
We next keep $u_1=u_2=u_0/\sqrt2$ fixed for
$\xi t_{\rm h}=100$ and switch off the atomic exchange.
For these post-loading diagnostics we use the passage clock
$\tau=t-T_\pi$, starting from $|e,g,0\rangle$. In the expressions
below, the argument $T$ denotes its endpoint $\tau=T$; hold and
retrieval times are measured relative to their respective onsets.
The nominal atomic-CD preparation yields $F_B(T)=0.99992176$, $F_{\Psi^+}(T)=0.99588074$, and $\mathcal C(T)=0.99587080$. The atomic weight of the exact target is
$Z=[1+q^2/2]^{-1}=0.99596634$. We note that most of the difference between the unconditional Bell
fidelity and unity comes from the intended photonic dressing, rather than imperfect atomic entanglement.

\begin{figure}[tbp]
\centering
\includegraphics[width=\textwidth]{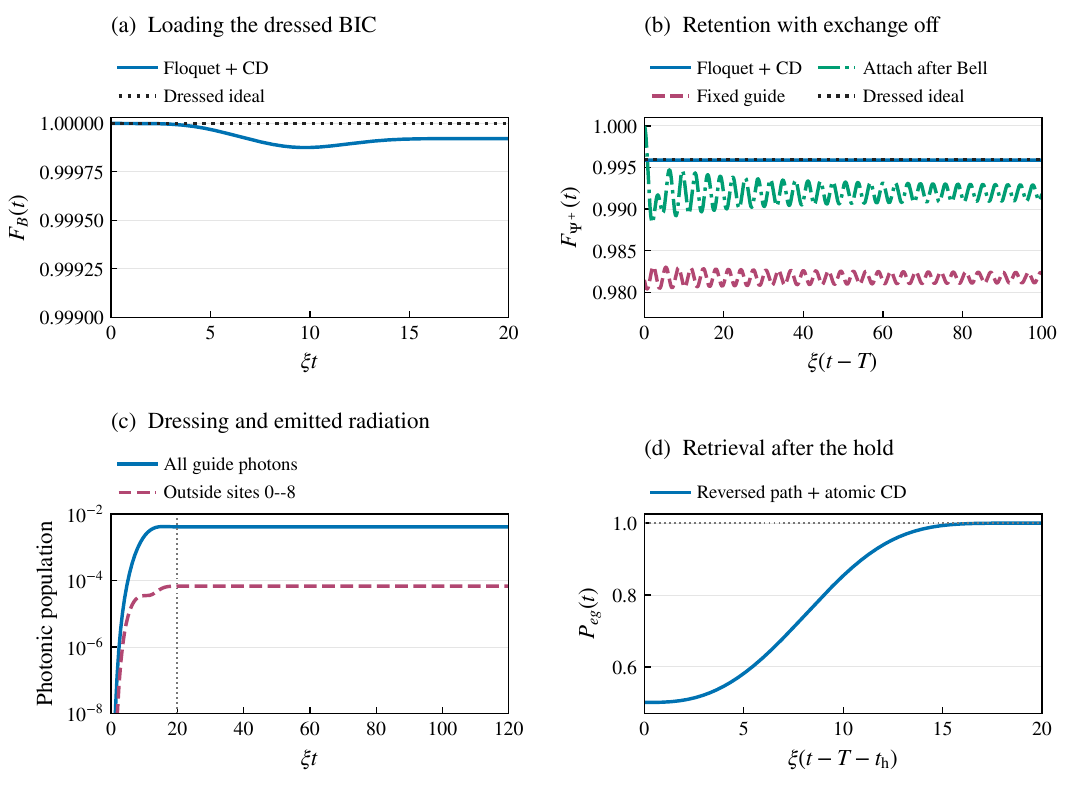}
\caption{Dressed-state loading, retention, and retrieval at $N_c=2004$,
$g/\xi=0.1$, $u_0=0.9$, and $\xi T=20$.
(a) Fidelity to the full compact BIC, comparing atomic CD with the ideal
dressed control. (b) Bell fidelity during a $\xi t_{\rm h}=100$ hold
with the exchange off, including fixed-guide and attach-after-Bell
references. (c) Total photonic population and population outside
sites $0,\ldots,8$ for the atomic-CD protocol. (d) Retrieval of the
atomic excitation by reversing the coupling path and atomic exchange
after the hold. In (a,c), the plotted time $t$ denotes the passage
clock $\tau$; the horizontal axes in (b,d) start at the beginning
of the hold and retrieval stages, respectively.}
\label{fig:dressed_retention}
\end{figure}

During the lossless effective hold, the overlap with the final BIC
is conserved because the target is an eigenstate of the
time-independent Hamiltonian. In figure~\ref{fig:dressed_retention}(a), we show the full-state
fidelity during preparation. In figure~\ref{fig:dressed_retention}(b),
the atomic Bell fidelity remains above $0.99588$ throughout the
sampled hold interval. By contrast,
preparing a vacuum-field Bell state with isolated exchange and then
attaching it to the final guide gives $F_B(T)=Z$ and
$F_{\Psi^+}(T+t_{\rm h})=0.99182$. At fixed guide couplings throughout
the same exchange pulse, the latter fidelity is $0.98226$.
These comparisons allow us to identify the effect of preparing the
photonic dressing together with the atomic state.

To distinguish the localized field from outgoing radiation, we
calculate the photonic probability outside sites $0,\ldots,8$
and compare it with the total waveguide population. This spatial
calculation confirms that the localized dressing and the emitted
field are distinct. Finally, reversing the coupling path and the sign of the atomic CD
for another $\xi T=20$ retrieves the excitation with
$P_{eg}=0.999864$. For context, the weak-drive example in
Ref.~\cite{PhysRevA.111.053711}, Fig.~4(a), reaches its first
Bell-fidelity maximum at $\xi t_{\max}=223$, with drive strength
$0.01\xi$ and $g=0.5\xi$. Our nominal effective sequence takes
$\xi(T_\pi+T)=22$, but uses an independent exchange and different
couplings. This contrast is not a resource-matched speedup;
stronger driving also shortens preparation in the reference protocol.

\subsection{Microscopic validation of the effective model}
\label{subsec:floquet_effective_validation}

To validate the central-sideband description used in the control protocol, we
compare Equation~\eqref{Hamiltonian}, supplemented by the laboratory-frame auxiliary exchange, with the corresponding effective Hamiltonian in a finite-mode
calculation. The comparison is performed during the passage stage,
starting from the locally prepared state \(|e,g,0\rangle\), and using the same
nominal parameters and the passage clock $\tau$. In the laboratory-frame simulation, the full
time-dependent frequencies \(\Omega_i(t)\) are retained, so all fast Floquet
oscillations are present in the numerical dynamics. The laboratory-frame state
is then transformed to the slow frame through
\(c_i^{\rm slow}(t)=e^{i\beta_i(t)\sin(\nu t)}c_i^{\rm lab}(t)\), before
comparing it with the central-sideband effective state. The counterdiabatic
coupling is included in both descriptions, with the corresponding micromotion
phases in the laboratory frame. We define the full-state agreement as
\begin{equation*}
F_{\rm state}(\tau)=
\left|\left\langle\Psi_{\rm eff}(\tau)
\middle|\Psi_{\rm micro}^{\rm slow}(\tau)\right\rangle\right|^2,
\end{equation*}
and $\Delta F_{\Psi^+}$ as the difference between the microscopic
slow-frame and effective atomic Bell fidelities. Here, $F_{\rm state}$ measures the agreement between the two dynamical
descriptions, while $F_B$ measures the fidelity to the selected
compact BIC.

\begin{figure}[tbp]
\centering
\includegraphics[width=\textwidth]{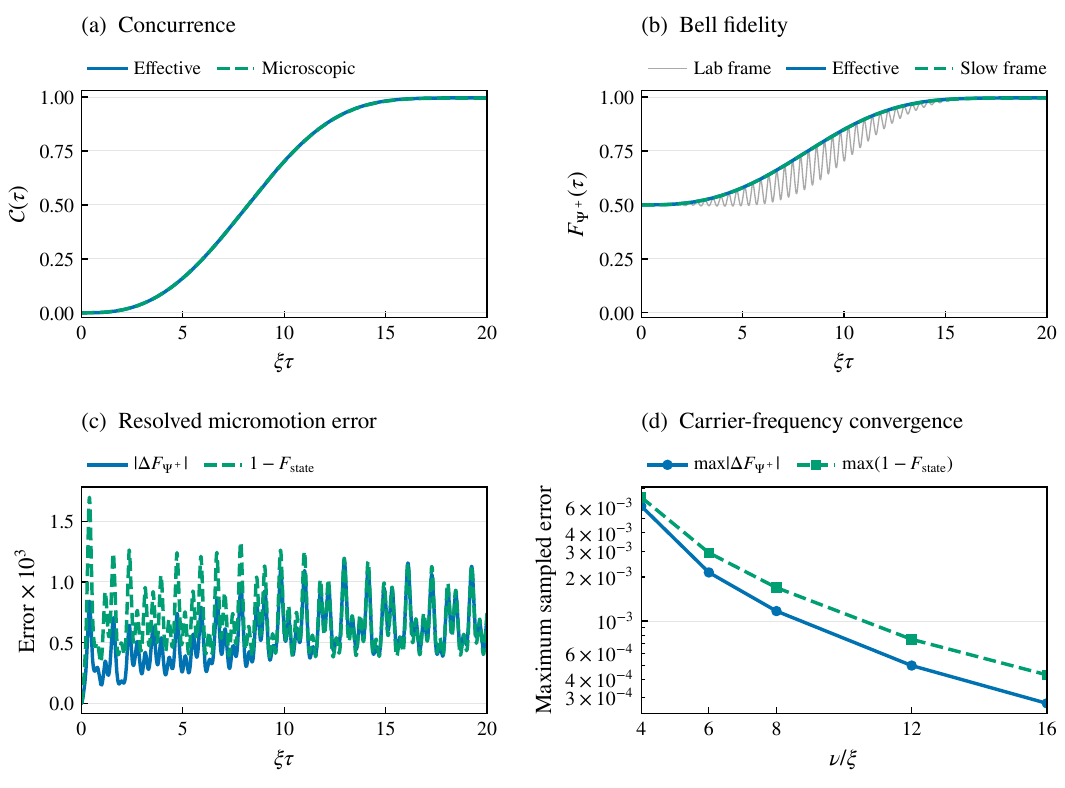}
\caption{Microscopic validation of the effective model.
(a) Concurrence during the passage, comparing the central-sideband
model with microscopic finite-mode dynamics after removal of
micromotion. (b) Bell fidelity: the gray curve is the raw
laboratory-frame result and the dashed curve is its slow-frame
counterpart. Concurrence is invariant under these local phase
rotations, whereas fidelity to a fixed Bell ket is not.
(c) Pointwise errors $|\Delta F_{\Psi^+}|$ and $1-F_{\rm state}$
at $\nu/\xi=8$, displayed in units of $10^{-3}$.
(d) Carrier-frequency convergence of the maximum sampled errors,
using 401 output times per frequency. In (a--c), $\tau$ is measured
from the start of the passage. Parameters are $N_c=2004$, $g/\xi=0.1$, $u_0=0.9$, and
$\xi T=20$; $\nu/\xi=8$ in (a--c).}
\label{fig:floquet_effective_validation}
\end{figure}

In figure~\ref{fig:floquet_effective_validation}, we compare the exact
laboratory-frame finite-mode dynamics with the effective model.
We observe good agreement throughout the Bell-BIC passage. For \(N_c=2004\), \(\nu/\xi=8\), \(u_0=0.9\), and
\(\xi T=20\), the maximum pointwise difference is
\(\max_t|\Delta F_{\Psi^+}|=1.17\times10^{-3}\) and the maximum slow-frame
state infidelity is \(1.70\times10^{-3}\). The final Bell fidelity is
\(F_{\Psi^+}=0.99588\) in the effective model and \(F_{\Psi^+}=0.99518\) in
the microscopic calculation after removal of micromotion.

We additionally vary $\nu/\xi$ over $4,6,8,12,16$ while holding the
effective path, duration, and exchange pulse fixed. The maximum sampled
Bell-fidelity error decreases from $6.09\times10^{-3}$ at $\nu/\xi=4$
to $2.75\times10^{-4}$ at $\nu/\xi=16$; the corresponding state
infidelity decreases from $6.99\times10^{-3}$ to $4.30\times10^{-4}$.
The fine time trace at $\nu/\xi=8$ resolves the fast oscillations;
the frequency sweep reports maxima on its specified output grid.
These results support the effective description of the passage for
the frequencies and durations considered here. The preceding loading,
hold, and retrieval calculations use the lossless effective model;
Sec.~\ref{sec:microscopic_reversible_cycle} presents a separate
microscopic write--hold--read test, starting after ideal loading.

\subsection{Control resources and implementation requirements}
\label{sec:cd_Hamiltonian}
The atomic auxiliary term is a two-body interaction,
\begin{equation}
H_{\rm cd}=i\dot\theta\left(
\sigma_2^+\sigma_1^--\sigma_1^+\sigma_2^-\right).
\end{equation}
It requires a phase-controlled exchange channel between the atoms,
in addition to their individual frequency drives. Experiments have demonstrated tunable complex hopping in superconducting
circuits~\cite{Roushan2017}, suggesting a possible implementation with
a parametric coupler. However, we do not assume a device-specific
realization of the present geometry. The local longitudinal frequency
drives alone do not generate this additional interaction.

Writing $\chi_j=\beta_j(t)\sin(\nu t)$, the laboratory-frame coefficient
of $\sigma_1^+\sigma_2^-$ must be
$-i\dot\theta\,e^{i(\chi_2-\chi_1)}$. The simulations include this phase,
which is necessary to obtain the desired exchange in the slow frame.
For the quintic ramp,
\begin{equation}
\max|\dot\theta|=\frac{15\pi}{32T}=0.0736311\,\xi
\qquad (\xi T=20).
\end{equation}
At the initial Bessel zero, $\beta_1=z_{01}=2.40482556$, the
in-phase modulation amplitude is $\nu z_{01}=19.2386\,\xi$ for
$\nu/\xi=8$. The envelope quadrature
$\dot\beta_j\sin(\nu t)$ is also required. An experimental implementation would therefore require checking
these amplitudes and phase-bandwidth requirements for the chosen device. The ideal dressed auxiliary in
Equation~\eqref{dressed_auxiliary} requires additional atom--field control.

\section{Preparation and retention under parameter errors}
\label{sec:k_sensitivity}
We now study how parameter errors affect the preparation and retention
of the entangled state. First, we consider phase matching, which
determines the stationary radiative cancellation of a BIC.
For a resonant wave vector $K=K_0+\delta K$, with $K_0=\pi/2$, we set
$\Omega(K)=\omega_c-2\xi\cos K$ and propagate the finite-mode Hamiltonian
in the frame rotating at this frequency. The same shift is applied to
every reference protocol. Away from the nominal point, a large final
Bell fidelity is not sufficient to identify the state as an exact BIC.

\begin{figure}[tbp]
\centering
\includegraphics[width=\textwidth]{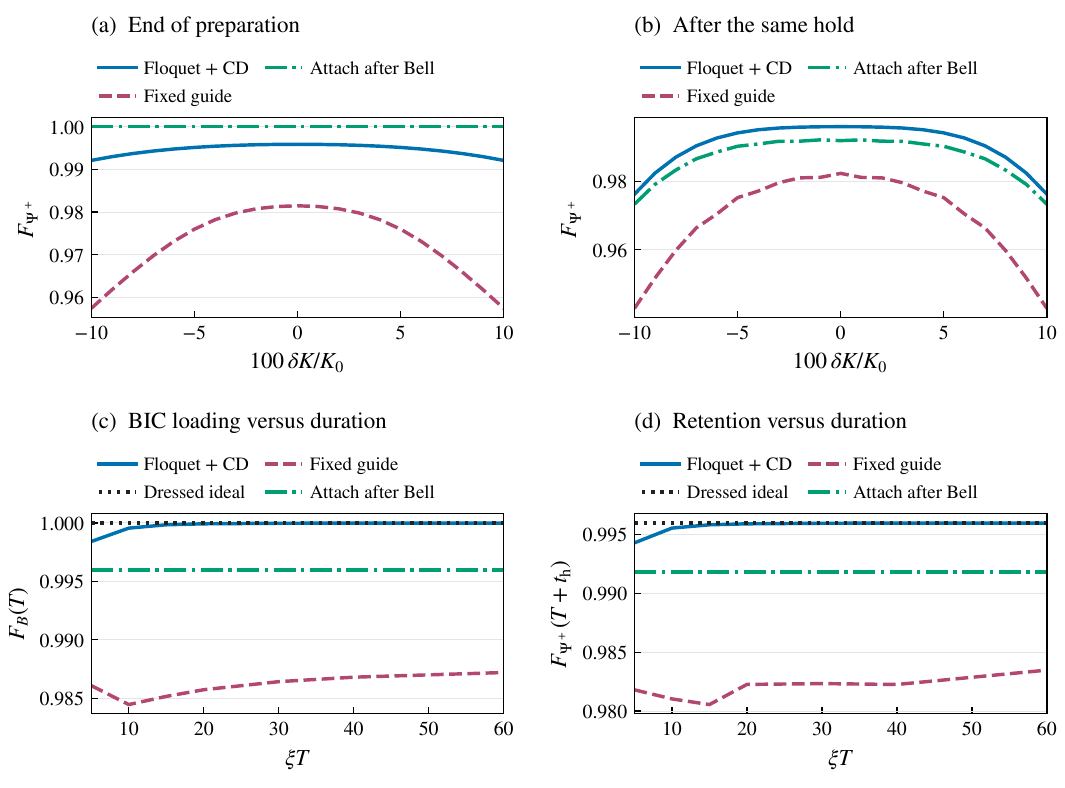}
\caption{Controlled comparisons of preparation and retention.
(a) Bell fidelity at the end of the passage and (b) after an identical
$\xi t_{\rm h}=100$ hold, as functions of wave-vector mismatch.
The geometry, bare coupling, initial state, duration $\xi T=20$, and
atomic exchange pulse are matched. \emph{Fixed guide} uses constant
$u_1=u_2=u_0/\sqrt2$; \emph{attach after Bell} uses isolated exchange and
a sudden connection to the final guide. (c) Full-state BIC fidelity
at the nominal $K_0$ versus passage duration, including ideal dressed
control. (d) Bell fidelity after the hold for the same durations.
All calculations use $N_c=2004$. The ideal dressed control and sudden
attachment are reference resources, not bandwidth-equivalent
experimental implementations.}
\label{fig:k_robustness_comparison}
\end{figure}

To compare the protocols in Fig.~\ref{fig:k_robustness_comparison},
we use the same connections $(0,6),(2,8)$, $g/\xi=0.1$, initial state $|e,g,0\rangle$,
passage duration, and atomic exchange pulse. The three practical
references are the proposed coupling path, fixed couplings
$u_1=u_2=u_0/\sqrt2$, and isolated exchange followed by attachment to the
same final guide. The first two have the same instantaneous coupling
norm $u_1^2+u_2^2=u_0^2$. Attachment is an idealized sudden switch,
so these are not bandwidth-matched hardware implementations.
All three use the same final Hamiltonian during a $\xi t_{\rm h}=100$
hold with the exchange off. The ideal dressed reference is included
only at the nominal geometry and requires the additional resource
specified in Equation~\eqref{dressed_auxiliary}.

At $\xi T=20$, the proposed protocol maintains
$F_{\Psi^+}(T)\geq0.99210$ across $|\delta K|/K_0\leq0.10$.
After the hold, its minimum over the same sampled wave-vector window
is $0.97620$, compared with $0.94289$ for fixed-guide exchange and
$0.97328$ for attachment after isolated Bell preparation.
These results show that a high preparation fidelity does not
necessarily imply equally robust retention. At the nominal point,
preparing the photonic dressing together with the atomic state reduces
the transient radiation that appears when a vacuum-field Bell state
is attached to the waveguide.

In the duration scan, we keep the exchange area fixed at $\pi/4$,
while the peak amplitude is proportional to $1/T$. This allows us
to study how fidelity depends on duration and pulse amplitude within
this control family. We compare the protocols under the conditions
specified above, without optimizing them under fixed amplitude and
bandwidth constraints. Appendix~\ref{app:markov} gives a separate static
Markovian reference for phase matching; the comparisons here use
the full finite-mode dynamics.

\subsection{Effective-control tolerances and consistency checks}
\label{subsec:protocol_parameter_robustness}

\begin{figure}[tbp]
\centering
\includegraphics[width=\textwidth]{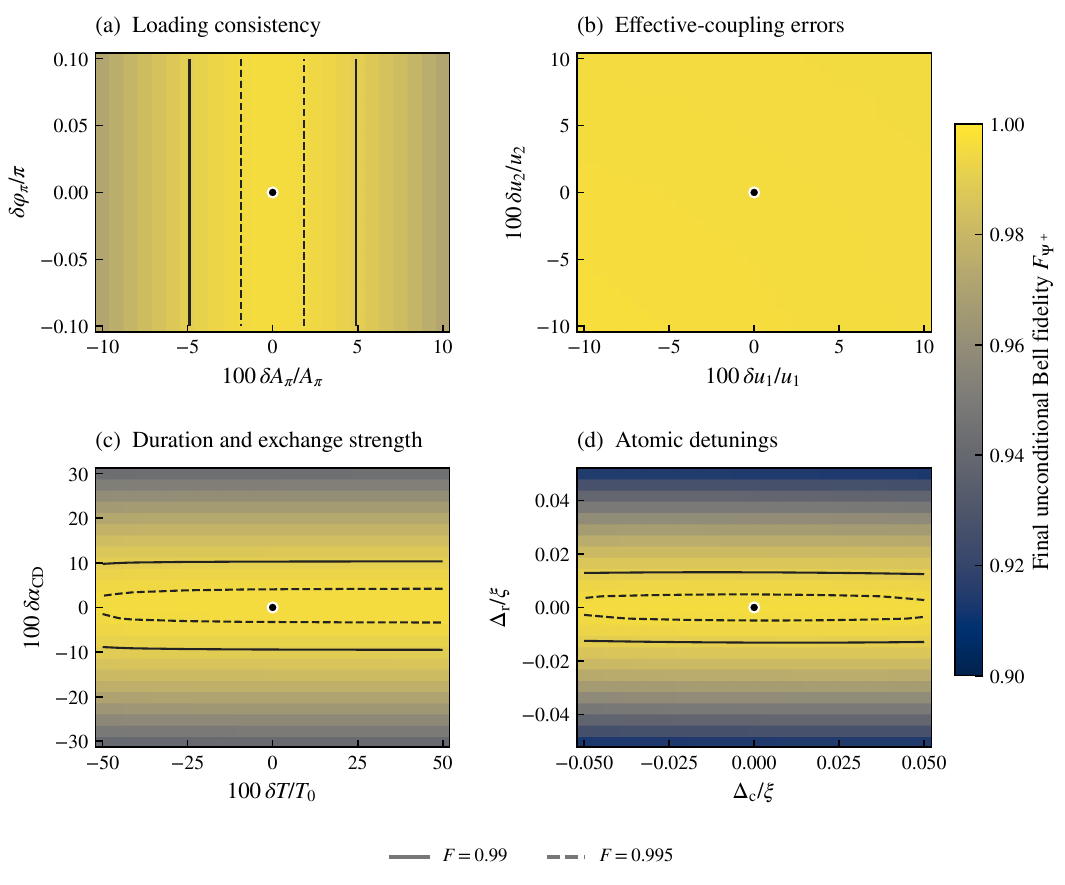}
\caption{Effective-model control tolerances at the end of the complete
two-stage preparation. (a) Loading area and phase; phase independence
is the expected symmetry of a single loading pulse. (b) Independent multiplicative errors
$\epsilon_{u_j}$, denoted $\delta u_j/u_j$ in the axes, which preserve
the zeros of $u_j$. (c) Duration errors $\delta T/T_0$ and fractional
exchange-amplitude errors $\delta\alpha_{\rm CD}$; the coupling
path and $\dot\theta$ are recalculated for each modified duration.
(d) Common and relative atomic detunings.
Solid and dashed contours mark $F_{\Psi^+}=0.99$ and $0.995$;
dots mark nominal settings. One common color scale is used without
smoothing the computed grid. Parameters are $N_c=804$,
$g/\xi=0.1$, $u_0=0.9$, $\xi T_\pi=2$, and $\xi T_0=20$.
These maps do not include a retention stage.}
\label{fig:protocol_parameter_robustness}
\end{figure}

In figure~\ref{fig:protocol_parameter_robustness}, we use the effective
model to calculate the unconditional Bell fidelity at the end of the
complete two-stage preparation. Here, each point corresponds to a
static calibration offset, rather than time-dependent noise. We parametrize the
multiplicative coupling errors by $\widetilde u_j(t)=(1+\epsilon_{u_j})u_j(t)$. This definition remains well-defined at a zero of $u_j$ and preserves
$u_1=0$ during loading. The axis notation $\delta u_j/u_j$ in
Fig.~\ref{fig:protocol_parameter_robustness}(b) denotes
$\epsilon_{u_j}$, rather than a pointwise division at those zeros.
Numerical settings are specified in the caption and
Appendix~\ref{app:numerics}.

For $|\epsilon_{u_j}|\leq0.10$, the minimum final fidelity exceeds
$0.995$. Relative atomic detuning is more restrictive: the sampled
final fidelities exceed $0.99$ for $|\Delta_{\rm r}|/\xi\leq0.0125$
over $|\Delta_{\rm c}|/\xi\leq0.05$, where
$\Delta_{1,2}=\Delta_{\rm c}\pm\Delta_{\rm r}/2$.
For the duration and CD-strength scan, we use
\begin{equation*}
\widetilde T=T_0+\delta T,\qquad
H_{\rm cd}^{\rm err}(\tau)
=(1+\delta\alpha_{\rm CD})\dot\theta_{\widetilde T}(\tau)\sigma_y.
\end{equation*}
Both the coupling schedule and its derivative are recalculated at
$\widetilde T$; $\delta\alpha_{\rm CD}$ then represents an
independent fractional exchange-amplitude error. The map therefore
measures sensitivity within the specified control family.

To understand the loading-phase scan, let us consider the role of
the pulse phase. Starting from $|g,g\rangle$, changing the phase of
the only loading pulse changes the common phase of the excited
branch. Since the subsequent dynamics conserves excitation number,
both concurrence and Bell fidelity remain unchanged. Therefore,
the flat direction in this scan is a consistency check of the
single-pulse loading symmetry; it does not test relative Bell-phase
errors. For a resonant pulse with fractional
area error $e$, the loaded population is
$\cos^2(\pi e/2)$, and the subsequent unconditional Bell fidelity is multiplied
by the same factor. It accounts for the observed minimum of $0.97151$
at $|e|=0.10$.

\subsection{Static calibration errors in modulation depths}
Until now, we have considered calibration errors in the effective
couplings. To study errors in the modulation depths themselves, we
introduce the following time-independent fractional offsets
\begin{equation}
\widetilde\beta_j(t)=(1+\epsilon_{\beta_j})\beta_j(t),\qquad
\widetilde u_j(t)=\mathcal J_0[\widetilde\beta_j(t)].
\label{physical_beta_errors}
\end{equation}
These errors shift the Bessel zeros and can leave nonzero effective
coupling where the nominal path specifies decoupling.
For this calculation, we start from the ideally loaded state
$|e,g,0\rangle$ and apply the errors only during the passage and the
subsequent hold. We keep the intended atomic exchange pulse unchanged.
In this way, we study the effective-model response to static errors
in the modulation depths without changing the exchange pulse or
the loading conditions.

\begin{figure}[tbp]
\centering
\includegraphics[width=\textwidth]{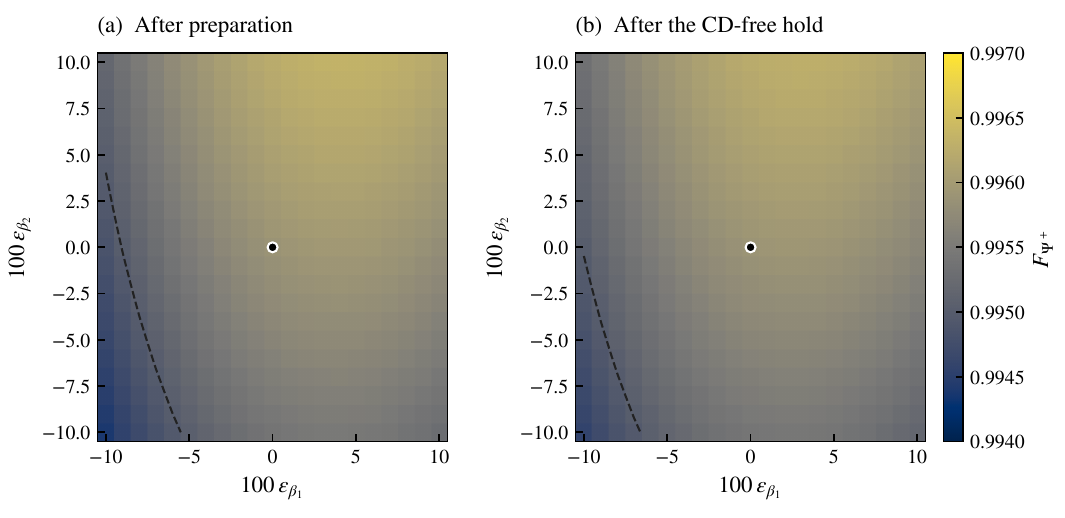}
\caption{Independent static fractional errors in the modulation depths,
applied after ideal local loading. (a) Unconditional Bell fidelity after
the $\xi T=20$ passage and (b) after a further $\xi t_{\rm h}=100$ hold
with atomic CD off. The perturbed couplings are
$\mathcal J_0[(1+\epsilon_{\beta_j})\beta_j(t)]$.
Both panels use the same color scale and a $21\times21$ grid at $N_c=804$.
Solid and dashed contours, where present, indicate $F_{\Psi^+}=0.99$
and $0.995$. Parameters otherwise match the nominal protocol.}
\label{fig:beta_errors}
\end{figure}

In figure~\ref{fig:beta_errors}, we compare preparation and retention
using a $21\times21$ grid of independent depth errors, with
$|\epsilon_{\beta_j}|\leq0.10$. We keep the perturbed final couplings
fixed for $\xi t_{\rm h}=100$ and switch off the exchange. Over this
grid, we obtain a minimum fidelity of $0.99439$ after preparation
and $0.99458$ at the end of the hold. Repeating the calculation at
the nominal point and the four corners gives agreement within
$10^{-10}$ for $N_c=804, 2004, 4004$. This calculation includes the displacement of the Bessel zeros,
which is absent from the multiplicative-$u_j$ map. Physically, the
offsets also modify the final couplings and the corresponding photonic
dressing. Therefore, a larger atomic Bell fidelity does not necessarily
mean a larger overlap with the nominal BIC target.

\subsection{Reversible preparation and a useful entanglement window}
\label{sec:microscopic_reversible_cycle}

We test a write--hold--read sequence starting from $|e,g,0\rangle$.
In this subsection, $t=0$ denotes the beginning of writing after
ideal loading; the loading interval is not included.
The coupling passage of duration $T$ is followed by a hold of duration
$t_h$ at the final couplings. Reading reverses the coupling envelopes and
the sign of the atomic exchange for a further time $T$, without reversing
the waveguide Hamiltonian. The return probability
$\eta_{\rm ret}(t_h)=\langle e,g|\rho_a(2T+t_h)|e,g\rangle$
quantifies recovery of this prepared excitation; it is not an arbitrary-state
quantum-memory fidelity. The intermediate atomic entanglement and the
overlap with the dressed target must be assessed separately.

We propagate the full time-dependent atom--guide Hamiltonian, retaining
the bare couplings and all Floquet harmonics, with independent external
relaxation and dephasing channels,
\begin{align}
 \dot\rho={}&-i[H_{\rm mic}(t),\rho]
  +\sum_j\left(\gamma_j\mathcal D[\sigma_j^-]
         +2\gamma_{\phi j}\mathcal D[\sigma_j^+\sigma_j^-]\right)\rho\nonumber\\
 &+\sum_x\left(\kappa\mathcal D[a_x]
        +2\kappa_\phi\mathcal D[a_x^\dagger a_x]\right)\rho,
 \label{eq:microscopic_cycle_lindblad}
\end{align}
where $\mathcal D[L]\rho=L\rho L^\dagger-\{L^\dagger L,\rho\}/2$,
$H_{\rm mic}=H(t)+H_{\rm cd}^{\rm lab}(t)$, and $a_x$ annihilates
a photon at guide site $x$. The relaxation rates are
$\gamma_j=1/T_{1,j}$ and $\kappa=1/T_{1,\gamma}$.
The explicitly represented waveguide is not counted again as an atomic
Markov decay channel. An exact local phase transformation is used for
numerical efficiency, without replacing the oscillating coupling phases
by their cycle averages. Relaxation feeds the vacuum, and dephasing is
retained in the full one-excitation density block. Observables are
unconditional; the block is never renormalized.
Specifically, the laboratory detuning is $\dot\chi_j(t)$, where
$\chi_j(t)=\beta_j(t)\sin(\nu t)$, including the envelope derivative.
The laboratory exchange has the matrix element
$\langle1|H_{\rm cd}^{\rm lab}|2\rangle
=-i\dot\theta\,e^{i(\chi_2-\chi_1)}$; its phase compensation is an
explicit auxiliary-control requirement, not a demonstrated property of
the waveguide-mediated exchange.

For a reference design, we use the atomic coherence times of device A in
Ref.~\cite{Kannan2020}, $T_{1,j}=(31.5,26.1)\,\mu\mathrm{s}$ and
$T_{2,j}^*=(4.2,3.6)\,\mu\mathrm{s}$, together with the lattice scales
$\xi/(2\pi)=6.25\,\mathrm{MHz}$, $T_{1,\gamma}=30\,\mu\mathrm{s}$,
and $T_{2,\gamma}^*\simeq3\,\mu\mathrm{s}$ of
Ref.~\cite{Ma2019Mott}. The design choices are
$g/(2\pi)=0.625\,\mathrm{MHz}$, $\nu/(2\pi)=50\,\mathrm{MHz}$,
$u_0=0.9$, and $\xi T=20$, giving $T=509.3\,\mathrm{ns}$.
The nominal peak modulation and exchange amplitudes, divided by $2\pi$,
are approximately $120.24$ and $0.460\,\mathrm{MHz}$, respectively.
We assume exponential, independent Markov dephasing with
$\gamma_{\phi j}=1/T_{2,j}^*-1/(2T_{1,j})$ and the analogous relation
for guide sites. Ramsey times alone do not establish this noise model,
and carrying these rates through the modulation is an additional assumption.
These component scales were measured in different devices; their combination,
the required geometry, and the auxiliary exchange are design assumptions.
The initialization pulse, readout errors, thermal excitation, frequency
disorder, and a calibrated noise spectrum are not included in this test.

The reference with fixed envelopes uses the same atomic exchange, duration,
and final couplings, isolating the benefit of shaping the couplings.
The additional unmodulated reference keeps the bare couplings throughout
writing and holding and therefore has a different final dressed target.
At the nominal write time, the shaped and fixed-envelope protocols yield
$F_{\Psi^+}(T)=0.92194$ and $0.90647$, respectively; the unmodulated
reference gives $0.87891$. After a $509.3\,\mathrm{ns}$ hold and the common
reverse readout, the return probabilities are $0.76418$ and $0.75162$
for the shaped and fixed-envelope preparations. Without external noise,
the shaped microscopic cycle returns the excitation with probability
$0.99906$. The improvement therefore concerns preparation and recovery
of the dressed excitation, not elimination of intrinsic decoherence.

We define the useful entanglement window by
\begin{equation}
 \tau_{\rm use}(F_{\rm req})=
 \sup\!\left\{h\geq0:\ F_{\Psi^+}(T+s)\geq F_{\rm req}
                  \ \text{for all }s\in[0,h]\right\},
 \label{eq:useful_bell_window}
\end{equation}
setting $\tau_{\rm use}=0$ if the prepared fidelity is already below the
threshold. For $F_{\rm req}=0.90$, the shaped and fixed-envelope windows
are approximately $91$ and $25\,\mathrm{ns}$; for $F_{\rm req}=0.85$ they
are $318$ and $248\,\mathrm{ns}$. Neither has a nonzero $0.95$ window
at the nominal duration. Reporting the threshold dependence avoids
interpreting a threshold-specific ratio as a universal lifetime gain.
Shorter passages mitigate dephasing, at the cost of increasing the peak
exchange according to $\max|\dot\theta|=15\pi/(32T)$; the duration scan
is not an optimization at fixed peak control strength.

The unconditional Bell fidelity can be obtained from three atomic
correlations without full tomography,
\begin{equation}
 F_{\Psi^+}=\frac{1+\langle\sigma_x^{(1)}\sigma_x^{(2)}\rangle
 +\langle\sigma_y^{(1)}\sigma_y^{(2)}\rangle
 -\langle\sigma_z^{(1)}\sigma_z^{(2)}\rangle}{4}.
 \label{eq:cycle_bell_correlations}
\end{equation}
We retain all trials, including relaxation events. For slow-frame
fidelities, the local tomography axes compensate for the known modulation
phases. These correlations characterize the reduced atomic state, not
the complete dressed BIC; the identity presumes two-level atoms, with
any higher-level leakage characterized separately.

\begin{figure}[tbp]
 \centering
 \includegraphics[width=\textwidth]{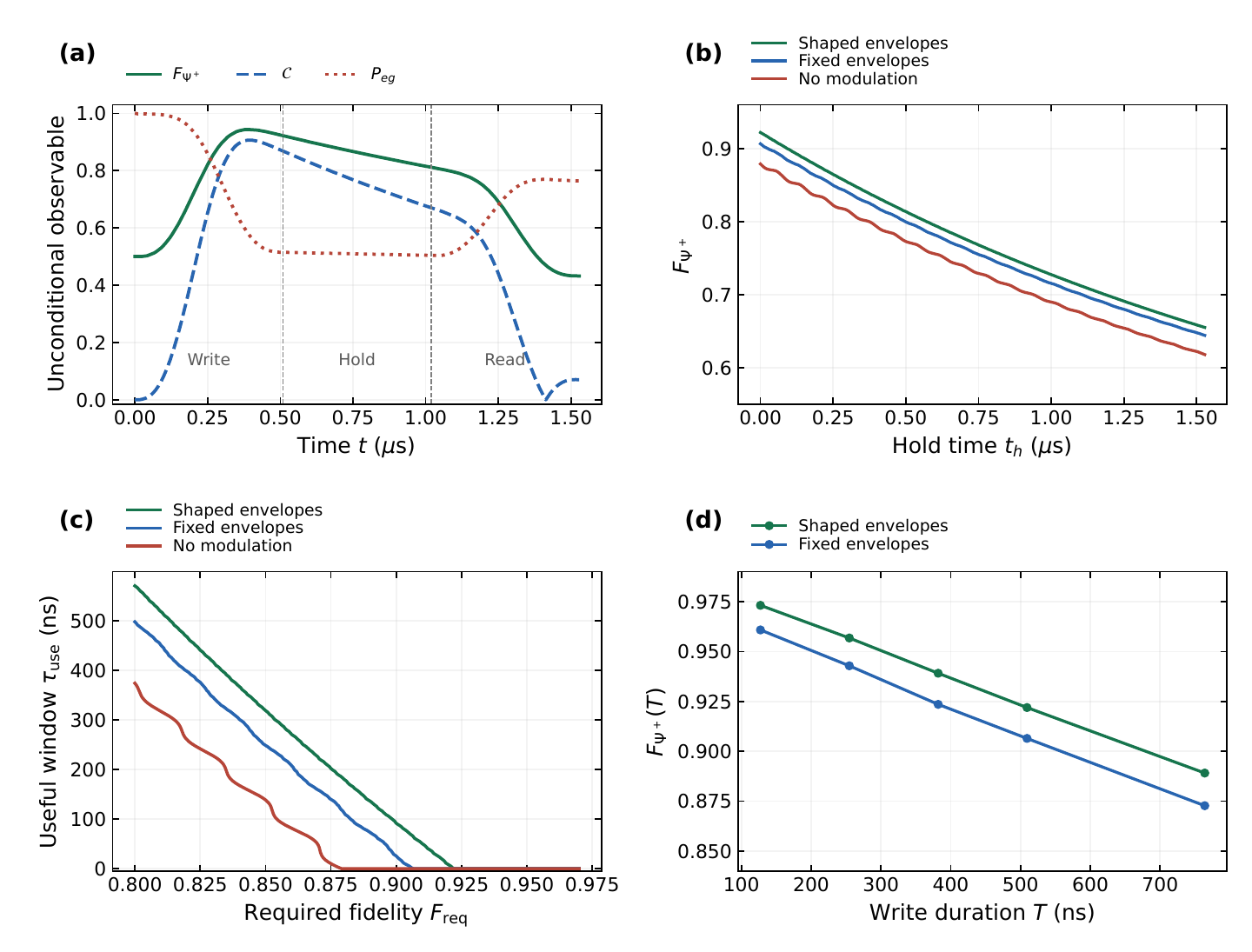}
 \caption{Microscopic feasibility test of reversible dressed-entanglement
 preparation with experimentally anchored coherence scales.
 (a) Unconditional Bell fidelity, concurrence, and atomic return population
 during writing, holding, and the reverse readout. Each stage lasts
 $509.3\,\mathrm{ns}$; dashed lines separate the stages.
 (b) Bell fidelity during a separate, extended hold after the same write
 pulse. The shaped-envelope and fixed-envelope preparations have the same
 atomic exchange and final couplings; the unmodulated reference retains
 the larger bare couplings and is not a matched-target ablation.
 (c) First-crossing useful window versus the required fidelity.
 (d) Write-time fidelity as a function of passage duration; both references
 use the same exchange waveform at each duration, but its peak changes
 from $\max|\dot\theta|/(2\pi)=1.84$ to $0.307\,\mathrm{MHz}$ across
 the scan. All curves retain the full microscopic modulation and the
 relaxation and dephasing channels in
 Eq.~\eqref{eq:microscopic_cycle_lindblad}. The atomic and guide parameters
 are drawn from different platforms, as specified in the text; this is
 not a reproduction of either experiment.}
 \label{fig:microscopic_reversible_bic}
\end{figure}
\section{Conclusions}

In this work, we have presented a counterdiabatic-assisted quasi-Floquet
protocol for preparing an entangled BIC from $|g,g,0\rangle$ in a
lossless effective waveguide model. We obtained an explicit compact eigenstate,
which allows us to distinguish conditional atomic entanglement,
unconditional Bell fidelity, and fidelity to the complete atom--field
state. Using this eigenstate, we derived a closed-form tracking residual
for an auxiliary control restricted to atomic exchange. This result
identifies the part of the photonic dressing evolution that the atomic
control does not reproduce, without requiring a spectral gap or
adiabatic selection of a unique state within the degenerate BIC sector.

The protocol starts with a local $\pi$ pulse that prepares
$|e,g,0\rangle$. Then, a smooth quasi-Floquet modulation changes the
atom--waveguide couplings through Bessel factors, defining a continuous
path toward the target BIC. An independent phase-controlled
counterdiabatic exchange suppresses atomic-rotation errors during the finite-time preparation. For $g/\xi=0.1$, $u_0=0.9$, and $\xi T=20$, our
numerical simulations of the effective model give a full-state BIC
fidelity of $F_B=0.99992$ and an unconditional atomic Bell fidelity of
$0.99588$. The difference between these fidelities reflects the
photonic component of the dressed BIC. By comparing the reference
protocols, holding the final couplings fixed with the exchange off,
and reversing the preparation path, we showed how preparing this
dressing contributes to the retention and recovery of the loaded
excitation. We also studied parameter errors, distinguishing their
effect on preparation from their effect on retention.

Using microscopic driven simulations, we tested the central-sideband
description during the preparation passage and validated the
quasi-Floquet effective model in the regime considered here. We also
simulated a microscopic write--hold--read cycle with relaxation and
dephasing, using experimentally motivated component scales. Shaping
the coupling envelopes increases the useful window with Bell
fidelity at least $0.90$ from approximately $25$ to $91\,\mathrm{ns}$
relative to fixed envelopes with the same exchange and final couplings.
This benefit is threshold- and noise-model-dependent. Microscopic
loading from the vacuum, exchange-phase errors, and device-specific
amplitude and bandwidth constraints remain to be assessed. The main
physical distinction is between generating atomic entanglement and
preparing the complete dressed BIC, which supports its retention after
the auxiliary exchange is removed. Our compact-state description
makes this distinction quantitative and identifies the tracking error
when the auxiliary interaction acts only on the atoms. We believe
that these results provide a useful basis for high-fidelity
preparation, retention, and retrieval of entangled excitations in
giant-atom waveguides, with tolerance to the static errors considered
in this work.

\data{
The lossless-model simulation code, configurations, and figure source data are
available through the versioned project repository in
Ref.~\cite{PRRCode2026}, at commit \texttt{3f4e844}. The reproducibility
material identifies the data and generators for the two-stage
preparation, dressed-state diagnostics, carrier validation, and
parameter scans. The additional microscopic dissipative-cycle code,
source data, and numerical checks are included as ancillary material
with this submission; they are not part of commit \texttt{3f4e844}.
}

\ack{
A.R.L. acknowledges support from ANID Anillo Project ATE250066.
P.A.O. acknowledges support from DGIIE USM PI-LIR-26-10 and
FONDECYT Grant No.~1230933. A.N. acknowledges support from FONDECYT
Regular Grant No.~1251131 and ANID Anillo Project ATE250066.
}

\appendix
\numberwithin{equation}{section}

\section{Geometric amplitude selection and validity of reduced descriptions}
\label{app:geometric_limits}

This appendix extends the geometric amplitude relation of
Ref.~\cite{Legon2026Geometric} to locally renormalized couplings.
We establish its amplitude and entanglement conditions for a specified
stationary family and assess the effects of neglecting separately
photonic dressing and freezing the control envelopes in the memory
integral. For controls held fixed at an independent value $t_0$, let
$u_i^0=\mathcal J_0[\beta_i(t_0)]$. For an initially empty guide,
the Laplace representation is
\begin{align}
\widetilde K_{ij}(s;t_0)
&=u_i^0u_j^0\sum_k\frac{g_{ik}g_{jk}^*}{s+i\omega_k},
\label{geo:laplace_kernel}\\
\widehat c(s;t_0)
&=\big[(s+i\Omega_0)I+\widetilde K(s;t_0)\big]^{-1}c(0).
\label{geo:laplace_solution}
\end{align}
Analytic continuation gives the retarded self-energy in
Equation~\eqref{controlled_selfenergy}. Here $t_0$ labels the frozen
stationary problem rather than the propagation time.

When $n_1=n_2$ and $u_1^0=u_2^0$, reciprocity of the full-zone guide
gives $\Sigma_{11}=\Sigma_{22}$ and $\Sigma_{12}=\Sigma_{21}$.
The Bell basis then diagonalizes the complete self-energy, including
its complex entries. In this symmetry limit,
\begin{align}
W_k^\pm&=\frac{u_1^0g_{1k}\pm u_2^0g_{2k}}{\sqrt2},
\label{geo:bell_channels}\\
\Sigma_\pm^R(E;t_0)
&=\sum_k\frac{|W_k^\pm|^2}{E-\omega_k+i0^+}.
\label{geo:scalar_selfenergy}
\end{align}
For a general normalized atomic direction $a$, the projected matrix
element is $W_k=a_1^*u_1^0g_{1k}+a_2^*u_2^0g_{2k}$. Its modulus
sets the projected radiative strength; a decoupled scalar eigen-channel
requires the additional symmetry conditions. Otherwise the full
matrix and the real-energy condition are retained.

Define the emission amplitude
\begin{equation}
\mathcal M(k)=\sum_j u_j^0g_{jk}^*a_j,\qquad
 g_{jk}^*=\frac{g}{\sqrt{N_c}}e^{ikx_j}(1+e^{ikn_j}).
\label{geo:emission_amplitude}
\end{equation}
At the geometric zeros $Kn_j=(2\ell_j+1)\pi$,
$\mathcal M(\pm K)=0$ for every atomic direction. Imposing
$\partial_k\mathcal M(\pm K)=0$ additionally selects a higher-order
zero. Since
\begin{equation}
\left.\partial_k g_{jk}^*\right|_K
=-\frac{ign_j}{\sqrt{N_c}}e^{iKx_j},
\label{geo:derivative_at_root}
\end{equation}
the positive root derivative condition gives
\begin{equation}
\frac{a_2}{a_1}=-\lambda_F e^{-iK\Delta x},\qquad
\lambda_F=\frac{n_1u_1^0}{n_2u_2^0}.
\label{geo:amplitude_ratio}
\end{equation}
For nonzero amplitudes and couplings, cancellation of the derivative
at the negative root also requires $e^{2iK\Delta x}=1$. The derivative
condition therefore selects a superposition within the radiatively
dark sector; it is an additional design condition, not a requirement
inferred from the retarded regulator.

An explicit family satisfying both the radiative and real-energy
conditions is
\begin{equation}
K=\pi/2,\quad n_j=4\ell_j+2,\quad\Delta x=2r,\quad
\Omega_0=\omega_c,
\label{geo:stationary_family}
\end{equation}
with integer lattice positions and nonnegative $\ell_j$. Two
unnormalized compact eigenstates are
\begin{equation}
|b_j\rangle=|j\rangle+
\frac{gu_j^0}{\xi}\sum_{l=1}^{n_j-1}
\sin\!\left(\frac{\pi l}{2}\right)|1_{x_j+l}\rangle.
\label{geo:compact_eigenstates}
\end{equation}
The field vanishes at all atomic connection sites, which belong to
the same lattice sub-lattice. Its hopping cancels the atomic sources,
so $(H_{\rm eff}-\omega_c)|b_j\rangle=0$. The normalized
combination selected by Equation~\eqref{geo:amplitude_ratio} has conditional
atomic direction
\begin{equation}
|a\rangle=\frac{|e,g\rangle-\lambda_F e^{-iK\Delta x}|g,e\rangle}
{\sqrt{1+\lambda_F^2}}.
\label{geo:atomic_direction}
\end{equation}
On the positive Bessel branch used here, $\lambda_F\geq0$, and
its concurrence and Bell fidelity are
\begin{align}
\mathcal C_a&=\frac{2\lambda_F}{1+\lambda_F^2},
\label{geo:conditional_concurrence}\\
F_{\Phi(\varphi)}^{(a)}
&=\frac{1-\mathcal C_a\cos(K\Delta x-\varphi)}{2},
\label{geo:conditional_bell_fidelity}
\end{align}
where $|\Phi(\varphi)\rangle=(|e,g\rangle+e^{-i\varphi}|g,e\rangle)/\sqrt2$.
Thus $u_1^0/u_2^0=n_2/n_1$ compensates unequal connection lengths
and produces a maximally entangled atomic component in this family.
A negative $\lambda_F$ introduces an additional relative phase of $\pi$.

Writing the full state as
\begin{equation}
|B\rangle=\sqrt Z\big(|a,0\rangle+|\phi_a\rangle\big),\qquad
Z=\frac{1}{1+\langle\phi_a|\phi_a\rangle},
\label{geo:dressed_state}
\end{equation}
the vacuum-field approximation is controlled by
$\chi=\langle\phi_a|\phi_a\rangle\ll1$, with exact error
\begin{equation}
1-|\langle a,0|B\rangle|^2=1-Z=\frac{\chi}{1+\chi}.
\label{geo:vacuum_error}
\end{equation}
The unconditional quantities are $\mathcal C=Z\mathcal C_a$ and
$F_\Phi=ZF_\Phi^{(a)}$. An exact normalized atomic direction can
therefore coexist with a non-negligible full-state error if the
photonic field is omitted.

An exactly photon-free special case occurs for identical, coincident
connection sets, $n_1=n_2$ and $\Delta x=0$. The ratio
$a_2=-u_1^0a_1/u_2^0$ then cancels $\mathcal M(k)$ at every momentum,
and $Z=1$ at any coupling strength. By contrast, the selected BIC in
the braided preparation geometry generally has localized photonic
dressing.

\begin{figure}[tbp]
\centering
\includegraphics[width=\textwidth]{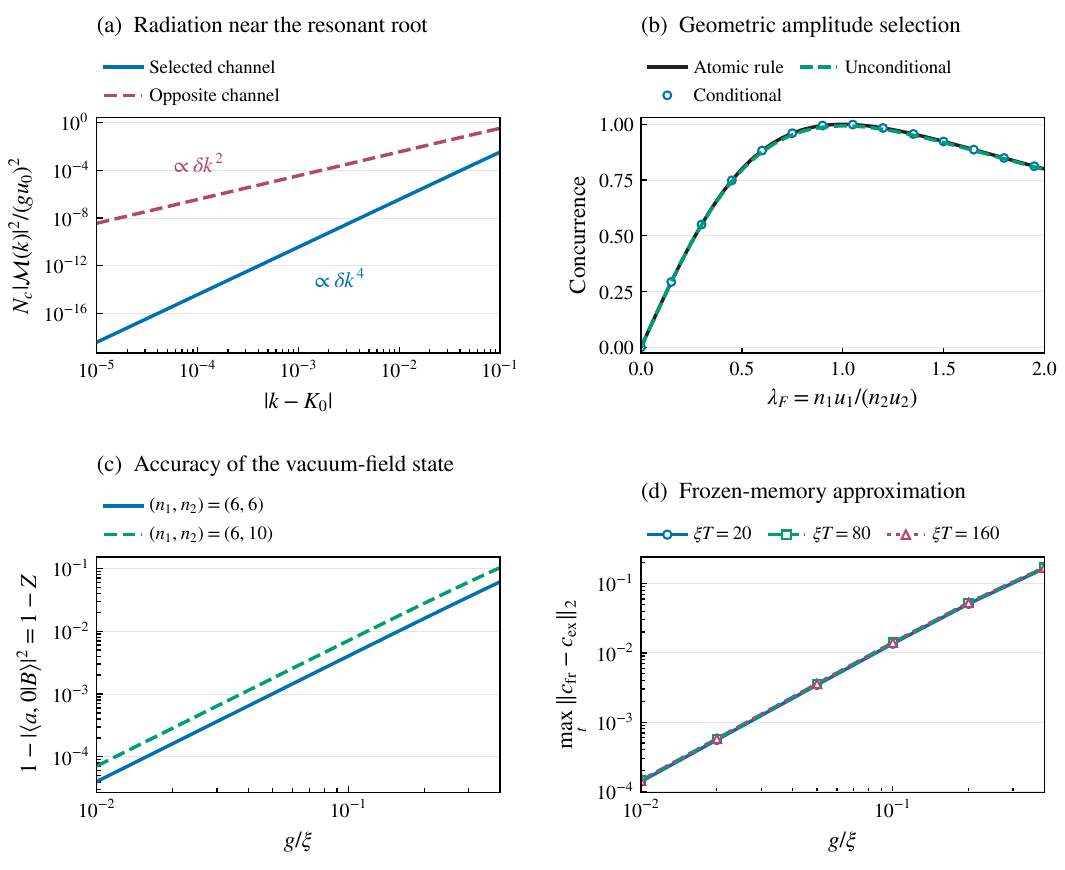}
\caption{Geometric amplitude selection and validity of reduced
descriptions. (a) Radiation intensity near $K_0=\pi/2$ for the
derivative-selected and opposite channels at $(n_1,n_2)=(6,6)$;
the same powers hold at the negative root. (b) Conditional and
unconditional concurrence of the exact dressed eigenstate for
$(n_1,n_2)=(6,10)$, compared with
Equation~\eqref{geo:conditional_concurrence}. (c) Full-state error from
neglecting the photon field at $\lambda_F=1$. (d) Atomic-amplitude
error from freezing the envelopes in the memory equation, compared
with exact unitary propagation for three passage durations. All
panels use $\Delta x=2$, $u_0=0.9$, and $N_c=2004$; panel (b)
uses $g/\xi=0.1$. The conditional identity and the two approximations
are evaluated separately.}
\label{fig:geometric_bic_approximation}
\end{figure}

Figure~\ref{fig:geometric_bic_approximation}(a--c) tests these
statements using the complete finite-mode Hamiltonian. For
$(n_1,n_2)=(6,6)$, $\Delta x=2$, and equal couplings, the selected
emission intensity scales as $\delta k^4$, whereas the opposite
atomic combination scales as $\delta k^2$ near either resonant root.
For $(6,10)$, varying $\lambda_F$ from zero to two reproduces the
conditional-concurrence relation in
Equation~\eqref{geo:conditional_concurrence}. Independent real-space and
momentum-space constructions agree within $8\times10^{-15}$, and
the largest eigenstate residual is below $2\times10^{-16}\xi$,
verifying the real-energy condition and radiative cancellation.
At $u_0^2=(u_1^0)^2+(u_2^0)^2=0.9^2$, $\lambda_F=1$, and
$\Delta x=2$, the two geometries give
\begin{equation}
\chi_{(6,6)}=\frac12\left(\frac{gu_0}{\xi}\right)^2,\qquad
\chi_{(6,10)}=\frac{15}{17}\left(\frac{gu_0}{\xi}\right)^2.
\label{geo:dressing_weights}
\end{equation}
For the nominal symmetric geometry at $g/\xi=0.1$, the
vacuum-state infidelity is $0.00403366$ and decreases quadratically
with $g$ in the weak-dressing limit. Unequal connection lengths
retain the atomic Bell condition but require the corresponding
geometry-dependent photonic normalization.

A separate dynamical approximation expands
\begin{align}
u_j(t-\tau)&=u_j(t)-\tau\dot u_j(t)+O(\tau^2),
\label{geo:envelope_expansion}\\
\dot u_j(t)&=-\mathcal J_1[\beta_j(t)]\dot\beta_j(t),
\label{geo:envelope_derivative}
\end{align}
and then replaces $u_j(t-\tau)$ by $u_j(t)$ inside the memory
integral. Here $\tau$ is the integration delay, distinct from the
passage clock used in Sec.~\ref{sec:two_stage_protocol}.
For varying envelopes, the replacement neglects
\begin{equation}
R_i(t)=\sum_j u_i(t)\int_0^t
[u_j(t-\tau)-u_j(t)]K_{ij}(\tau)c_j(t-\tau)\,d\tau.
\label{geo:frozen_residual}
\end{equation}
A criterion based on $|u_j/\dot u_j|$ requires nonzero $u_j$ and
an established reservoir memory scale. Since the tight-binding
kernel has long-time tails, and the carrier frequency alone does not
provide a memory cutoff, we assess this approximation directly. The frozen-envelope equation admits an auxiliary-ODE representation,
\begin{align}
\dot p_{jk}&=-i\omega_k p_{jk}-ig_{jk}^*c_j^{\rm fr},
\label{geo:memory_ode}\\
\dot c_i^{\rm fr}&=-i\sum_j(H_a)_{ij}c_j^{\rm fr}
-i u_i(t)\sum_k g_{ik}\sum_j u_j(t)p_{jk}.
\label{geo:frozen_atomic_ode}
\end{align}
Here $p_{jk}(0)=0$ and $H_a$ is the atomic Hamiltonian, including
the bare atomic energy and the same atomic CD as in the
exact comparison. The $p_{jk}$ are memory variables rather than
independent photon modes, and the atomic amplitudes are not
renormalized. For constant $u_j$, reconstructing
$\phi_k=\sum_j u_jp_{jk}$ recovers the complete vacuum-input
dynamics, as checked numerically. For the quintic passage from $|e,g,0\rangle$, we compare
Equation~\eqref{geo:frozen_atomic_ode} with the unitary finite-mode
propagation using
\begin{equation}
\epsilon_c=\max_{0\leq t\leq T}
\norm{c^{\rm fr}(t)-c^{\rm ex}(t)}_2.
\label{geo:atomic_error_metric}
\end{equation}
Figure~\ref{fig:geometric_bic_approximation}(d) shows the controlled
weak-coupling limit: at $g/\xi=0.01$ and $\xi T=20$,
$\epsilon_c=1.39\times10^{-4}$, compared with $1.36\times10^{-2}$
at $g/\xi=0.1$. At the latter coupling, increasing $\xi T$ to $80$
and $160$ gives $1.40\times10^{-2}$ and $1.41\times10^{-2}$,
respectively. Slowing the complete path does not therefore remove
the accumulated frozen-envelope error. At $g/\xi=0.1$ and
$\xi T=20$, the maximum difference in the Bell-overlap expression
is $3.50\times10^{-3}$, which is appreciable on the scale of the
main preparation error. At larger couplings, the approximate atomic
norm can exceed unity, so these overlaps cannot generally be
interpreted as fidelities of a physical reduced state.

All plotted dynamical comparisons use $N_c=2004$, relative and
absolute tolerances $10^{-10}$ and $10^{-12}$, maximum step
$0.04/\xi$, and 801 output times. The longest interval is $160/\xi$.
At $g/\xi=0.1,0.4$ and $\xi T=20,160$, repeating the calculation
with $N_c=804,2004,4004$, tolerances $10^{-11},10^{-13}$, and
maximum step $0.02/\xi$ changes $\epsilon_c$ by less than
$3\times10^{-13}$.

The stationary amplitude relation, weak-dressing approximation, and
frozen-envelope approximation thus have distinct validity conditions.
The preparation protocol in the main text retains the full
finite-mode dynamics without envelope freezing.

\section{Numerical propagation and consistency checks}
\label{app:numerics}
For the lossless amplitude calculations, we discretize the full Brillouin zone as
$k_m=-\pi+2\pi m/N_c$, $m=0,\ldots,N_c-1$, with periodic boundary
conditions in the corresponding real-space guide.
The basis for the passage is
$(c_1,c_2,\phi_0,\ldots,\phi_{N_c-1})$; the two-stage preparation adds
the vacuum amplitude $c_0$. The normalization
$g_{jk}\propto N_c^{-1/2}$ is retained when changing grid size.

We propagate the atomic and photonic amplitudes using the following
equations, without constructing the full Hamiltonian matrix:
\begin{align}
\dot c_i&=-i\left[\Delta_i c_i+\sum_k f_{ik}(t)\phi_k
+\sum_j(H_{\rm aux})_{ij}c_j\right],\\
\dot\phi_k&=-i\left[(\omega_k-\Omega_0)\phi_k+
\sum_j f_{jk}^*(t)c_j\right].
\end{align}
These equations describe the atomic-CD implementation. For the ideal
dressed reference, we additionally apply $-iH_{\rm dress}|\Psi\rangle$
to the full atom--field right-hand side. The local loading matrix
elements are included when the drive is on.
MATLAB's variable-order Adams solver \texttt{ode113} propagates these
equations without Markovian elimination. The dressed-state and
matched-reference calculations use relative and absolute tolerances
$10^{-10}$ and $10^{-12}$ and maximum step $0.04/\xi$.
The two-stage trajectory and effective-control maps use
tolerances $(10^{-9},10^{-11})$ and
$(10^{-8},10^{-10})$. The physical-depth maps use
$(10^{-9},10^{-11})$ and maximum step $0.05/\xi$.

For the vacuum-plus-one-excitation state
$c_0|G\rangle+c_1|1\rangle+c_2|2\rangle+\sum_k\phi_k|k\rangle$,
the unconditional atomic observables are
\begin{align}
P_a&=|c_1|^2+|c_2|^2,\nonumber\\
\mathcal C&=2|c_1c_2^*|,\qquad
F_{\Psi^+}=\frac{|c_1+c_2|^2}{2}.
\label{dynamic_atomic_observables}
\end{align}
The concurrence formula follows from the absence of $|e,e\rangle$
population~\cite{Wootters1998}, including during coherent loading.
For $P_a>0$, conditioning with
$P_{1a}=|e,g\rangle\langle e,g|+|g,e\rangle\langle g,e|$ gives
$\rho_a^{\rm cond}=P_{1a}\rho_aP_{1a}/P_a$,
$\mathcal C_{\rm cond}=\mathcal C/P_a$, and
$F_{\Psi^+}^{(a)}=F_{\Psi^+}/P_a$.
These conditional quantities are undefined at $P_a=0$.
During loading, conditioning on one atomic excitation differs from
conditioning only on the photon vacuum, which also retains $c_0$.
All pure-target fidelities use the squared-overlap convention.

To obtain Equation~\eqref{duhamel_bound}, choose the phase of
$|\Phi(t)\rangle=e^{i\alpha(t)}|B(t)\rangle$ so that the component
of $(i\partial_t-H)|\Phi\rangle$ along $|\Phi\rangle$ vanishes,
where $H=H_{\rm eff}+H_{\rm aux}$. The remaining defect is
$e^{i\alpha}|r\rangle$. For the same initial state, Duhamel's identity
and unitary propagation give
\begin{equation}
\norm{\Psi(T)-\Phi(T)}\leq\int_0^T\norm{r(t)}\,dt.
\end{equation}
For normalized states,
$1-|\langle\Phi|\Psi\rangle|^2\leq\norm{\Psi-\Phi}^2$,
which proves the stated infidelity bound.

The instantaneous target is evaluated from Equation~\eqref{compact_bic},
rather than extracted from a single numerically selected eigenvector
in a degenerate eigenspace. We check its norm, the residual
$(H_{\rm eff}-E_B)|B\rangle$, its analytic parameter derivative, and
the action of both atomic and dressed auxiliary controls.
The largest nominal eigenstate residual is below $10^{-15}\xi$.
For $N_c=804,2004,4004$, the nominal preparation and hold fidelities
agree within $10^{-10}$. These calculations verify the finite-grid convergence of the propagated
observables. The tracking analysis does not rely on a finite-size gap.

The real-space photon amplitudes are obtained by the unitary discrete
Fourier transform. Outgoing population is summed outside sites
$0,\ldots,8$. The maximum group velocity is $2\xi$, so propagation to
the opposite boundary of the smallest guide takes approximately
$N_c/(4\xi)=201/\xi$, longer than the longest nominal sequence
$T+t_{\rm h}+T=140/\xi$ and the largest duration-scan interval
$160/\xi$. Grid convergence provides an additional check on boundary
effects. No phenomenological waveguide dissipator is added in these
lossless calculations. The external losses of
Sec.~\ref{sec:microscopic_reversible_cycle} are treated separately below.

For the physical-depth maps, the nominal inverse-Bessel trajectory is
tabulated at 4001 equally spaced times and interpolated with a
shape-preserving cubic interpolant; actual couplings are evaluated
as Bessel functions of the perturbed depths. The inverse is taken on the first monotonic branch,
$0\leq\beta_j\leq z_{01}$, with $\mathcal J_0(z_{01})=0$; the
two-stage waveforms use a solver bracketed on this interval. For microscopic
passage tests we retain both modulation quadratures and transform
the atomic amplitudes by
$c_i^{\rm slow}=e^{i\beta_i(t)\sin\nu t}c_i^{\rm lab}$.
Frequency-sweep maxima use 401 output times, with internal steps
bounded by both $0.01/\xi$ and one-fiftieth of the carrier period.
The fine $\nu/\xi=8$ traces use a denser output grid.
Full final states, nominal trajectories, solver settings, and source
tables are retained in the reproducibility material cited in
Ref.~\cite{PRRCode2026}.

For the dissipative microscopic cycle in
Sec.~\ref{sec:microscopic_reversible_cycle}, we instead propagate
the full one-excitation density block $R$ in real space. With no
loading drive or thermal excitation, the vacuum is absorbing and
its population is $1-\Tr R$. Writing $h$ for the microscopic
Hamiltonian in the nominal local-phase frame, the block obeys
\begin{equation}
 \dot R_{ab}=-i[h,R]_{ab}
 -\frac{\Gamma_a+\Gamma_b}{2}R_{ab}
 -(1-\delta_{ab})(\varphi_a+\varphi_b)R_{ab},
\end{equation}
where $\Gamma_a=(\gamma_1,\gamma_2,\kappa,\ldots)$ and
$\varphi_a=(\gamma_{\phi1},\gamma_{\phi2},\kappa_\phi,\ldots)$.
The phase factors $g e^{i\chi_j(t)}$ are retained without cycle
averaging. A fourth-order Runge--Kutta solver uses $\xi\,dt=0.01$
and stores observables every $0.05/\xi$. The cycle uses $N_c=192$
guide sites and the extended hold uses $N_c=256$; these are
numerical grids, not fabricated device sizes. Halving the time step
changes the tested nominal observables by less than $3\times10^{-9}$;
increasing the respective guide sizes to $256$ and $384$ changes
them by less than $2\times10^{-15}$. An independent laboratory-frame
Lindblad calculation, including vacuum recycling, checks the
transformation and dissipators on a smaller grid.
For this mixed-state evolution,
$\mathcal C=2|R_{12}|$ and
$F_{\Psi^+}=(R_{11}+R_{22})/2+\operatorname{Re}R_{12}$;
the pure-amplitude formulas above are not used. Density-matrix
Hermiticity, trace, final positivity within numerical tolerance,
and the three-correlator identity are checked without renormalization.

\section{State-selective derivation of the counterdiabatic control}
\label{app:restricted_cd_derivation}

Let $H_0(t)$ denote the complete atom--waveguide Hamiltonian in the
central-sideband model during the passage, without the loading pulse
or auxiliary exchange. At the nominal parameters, the compact state
satisfies $H_0(t)|B(t)\rangle=E_B|B(t)\rangle$. The reference for
counterdiabatic driving is $H_0$, not the Hamiltonian after the auxiliary
term has been added. All statements of exact tracking below refer to
this effective model, rather than to the laboratory-frame driven
Hamiltonian before the sideband approximation.

\paragraph{Exact tracking of the selected eigenstate.}
For the smooth rank-one projector $P_B=|B\rangle\langle B|$, define
\begin{equation}
K_B=i[\dot P_B,P_B].
\label{eq:restricted_projector_generator}
\end{equation}
This is the state-selective form of transitionless
driving~\cite{Berry2009}. Differentiating $P_B^2=P_B$ gives
\begin{equation*}
P_B\dot P_BP_B=0,\qquad [[\dot P_B,P_B],P_B]=\dot P_B.
\end{equation*}
Since $[H_0,P_B]=0$, it follows that
\begin{equation}
i\dot P_B=[H_0+K_B,P_B].
\end{equation}
Thus a state initially in the range of $P_B(0)$ follows $P_B(t)$
exactly under $H_0+K_B$, up to phase. This construction uses the
known smooth eigenstate and does not divide by energy differences.
It therefore remains well defined for the selected compact BIC even
though its eigenvalue is embedded and degenerate.

Degeneracy nevertheless matters: $P_B$ specifies a particular
one-dimensional path inside the degenerate eigenspace. Preserving
that entire eigenspace does not fix this path. Consequently, this is
an explicitly state-selective construction, not a claim that the
uncontrolled adiabatic evolution uniquely selects $|B(t)\rangle$, or
that $K_B$ transports every other instantaneous eigenstate.

\paragraph{Tangent directions of the compact BIC.}
For fixed $q=gu_0/\xi$, write
\begin{align}
|B\rangle&=\cos\eta\,|D\rangle+\sin\eta\,|p\rangle,
&\tan\eta&=\frac{q}{\sqrt2}\sin(2\theta),\\
|D\rangle&=\cos\theta|1\rangle+\sin\theta|2\rangle,
&|p\rangle&=\frac{|1_1\rangle+|1_{n+1}\rangle}{\sqrt2}.
\end{align}
Here $|1\rangle,|2\rangle$ include the photonic vacuum and
$|p\rangle$ includes both atoms in their ground states. Define
$|D_\perp\rangle=-\sin\theta|1\rangle+\cos\theta|2\rangle$ and
$|R\rangle=-\sin\eta|D\rangle+\cos\eta|p\rangle$.
The vectors $|D_\perp\rangle$ and $|R\rangle$ are mutually
orthonormal and orthogonal to $|B\rangle$. In this real gauge,
$\langle B|\dot B\rangle=0$, and
\begin{align}
|\dot B\rangle&=\cos\eta\,\dot\theta|D_\perp\rangle
 +\dot\eta|R\rangle,\\
\dot\eta&=\frac{\sqrt2 q\cos(2\theta)}
 {1+(q^2/2)\sin^2(2\theta)}\dot\theta.
\end{align}
The first direction rotates the normalized atomic component; the
second changes the atom--photon dressing. Both follow directly from
the instantaneous eigenstate of the complete $H_0$.

\paragraph{Optimal instantaneous control within the exchange ansatz.}
Restrict the available auxiliary Hamiltonian to
$H_a(t)=a(t)\sigma_y$, where $a(t)$ is real and
$\sigma_y=i(|2\rangle\langle1|-|1\rangle\langle2|)$ acts as zero
on the photonic subspace. Motivated by constrained
counterdiabatic control~\cite{SelsPolkovnikov2017}, we minimize the
state-specific instantaneous tracking defect
\begin{equation}
\mathcal L(a)=\left\|(1-P_B)
\left(i|\dot B\rangle-a\sigma_y|B\rangle\right)\right\|^2.
\label{eq:restricted_tracking_cost}
\end{equation}
This is a selected-state objective, not the Hilbert--Schmidt action
averaged over the full spectrum. Since
$\sigma_y|D\rangle=i|D_\perp\rangle$ and $\sigma_y|p\rangle=0$,
\begin{align}
|r_a\rangle&=i\cos\eta(\dot\theta-a)|D_\perp\rangle
 +i\dot\eta|R\rangle,\\
\mathcal L(a)&=Z(\dot\theta-a)^2+\dot\eta^2,
\qquad Z=\cos^2\eta.
\label{eq:restricted_cost_evaluated}
\end{align}
For $Z>0$, the unique minimizing coefficient is
\begin{equation}
a_{\rm opt}=\dot\theta,\qquad
H_{\rm cd}^{(a)}=\dot\theta\sigma_y,\qquad
\min_a\mathcal L(a)=\dot\eta^2.
\label{eq:restricted_cd_optimum}
\end{equation}
The coefficient is therefore fixed by the tangent of the selected
full dressed eigenstate, rather than by a freely chosen Bell-pulse
area. Its optimality is instantaneous and restricted to this control
ansatz; it does not imply globally optimal final fidelity, minimum
protocol time, or exact tracking of the dressed state.

\paragraph{Exact completion and physical limitation.}
An exact state-selective completion is
\begin{equation}
H_{\rm dress}=i\dot\eta
\left(|p\rangle\langle D|-|D\rangle\langle p|\right),
\end{equation}
because
\begin{equation}
\left(H_{\rm cd}^{(a)}+H_{\rm dress}\right)|B\rangle
=i|\dot B\rangle.
\end{equation}
The sum has the same action on the target as $K_B$, although the
two operators need not agree on its orthogonal complement. In
particular, the literal atomic block of $K_B$ is
$\Pi_aK_B\Pi_a=Z\dot\theta\sigma_y$, with
$\Pi_a=|1\rangle\langle1|+|2\rangle\langle2|$.
Thus $\dot\theta\sigma_y$ should not be described as merely the
atomic block of the projector generator. It is the minimizer of
the restricted state-tracking objective above.

The need for an additional resource also follows from photon-number
balance. Let $\Pi_\gamma$ project onto the photonic sector. On an
instantaneous eigenstate of $H_0$, the expectation value of
$[H_0,\Pi_\gamma]$ vanishes; any purely atomic exchange commutes with
$\Pi_\gamma$. Hence $H_0+H_a$ gives zero instantaneous derivative of
the photonic population when evaluated on $|B\rangle$, whereas
the target requires
$d\langle\Pi_\gamma\rangle_B/dt=2\sin\eta\cos\eta\dot\eta$.
When this quantity is nonzero, exact dressed-state tracking is
impossible with atomic exchange alone. The physical dynamics may
still build the required dressing approximately by departing
slightly from the instantaneous eigenstate.

$H_{\rm dress}$ requires an additional controlled atom--photon
interaction and is not supplied automatically by the longitudinal
modulation. It is therefore an exact theoretical benchmark.
The atomic exchange remains an entangling resource even at $g=0$;
the contribution of the guide-coupling protocol must be established
through full-state BIC fidelity and subsequent retention, not
through Bell generation alone.

\section{A trace-preserving static reference}
\label{app:markov}
To compare with static phase-matching descriptions, we introduce a
Markovian reference using the same tight-binding conventions. For identical connection separation $n$, the tight-binding
on-shell Green function gives
\begin{align}
A_{ij}(K)&=\frac{g^2}{2\xi\sin K}\big[
2e^{iK|x_i-x_j|}\nonumber\\
&\quad+e^{iK|x_i+n-x_j|}
+e^{iK|x_i-x_j-n|}\big],\quad 0<K<\pi.
\label{eq:pra_collective_matrix}
\end{align}
Here $J_{ij}=\operatorname{Im}A_{ij}$ and
$\gamma_{ij}=\operatorname{Re}A_{ij}$. The master equation is
\begin{align}
\dot\rho&=-i[H_a+H_{\rm ex},\rho]+\sum_{ij}\gamma_{ij}\mathcal D_{ij}[\rho],\\
\mathcal D_{ij}[\rho]&=2\sigma_i^-\rho\sigma_j^+
-\{\sigma_j^+\sigma_i^-,\rho\},\\
H_{\rm ex}&=\sum_{ij}J_{ij}\sigma_i^+\sigma_j^- .
\end{align}
Here $H_a=\Omega\sum_i\sigma_i^+\sigma_i^-$ denotes the bare
atomic Hamiltonian, or its rotating-frame counterpart.
The sum over both indices includes the Hermitian-conjugate
exchange terms, and $1/\sin K$ retains the tight-binding
density-of-states factor. We check trace preservation, ground-state stationarity, Hermiticity,
and positivity in the numerical implementation. This static reference
is not used in the finite-mode propagation of the main results.

\bibliographystyle{qst_num}
\bibliography{Refs}
\end{document}